\documentstyle[prb,preprint,aps]{revtex}
\font\ee=msbm10 scaled \magstep1
\font\ee=msbm10 scaled \magstep1

\begin{document}
\title
{\bf \Large Linear Canonical Transformations and Quantum Phase: 
A unified canonical and algebraic approach} 
\author{T. Hakio\u{g}lu}
\address{Physics Department,
Bilkent University, Ankara 06533 Turkey}
\maketitle
\begin{abstract}
The algebra of generalized linear quantum canonical transformations 
is examined in the perspective of Schwinger's unitary-canonical operator 
basis. Formulation of the quantum phase problem within the theory of quantum 
canonical transformations and in particular with the generalized 
quantum action-angle phase space formalism is established and it is 
shown that the conceptual foundation of the quantum phase problem lies 
within the algebraic properties of the canonical transformations 
in the quantum phase space. 
The representations of the Wigner function in the generalized 
action-angle unitary operator pair for certain Hamiltonian systems with 
dynamical symmetry are examined. This generalized canonical formalism is 
applied to the quantum harmonic oscillator to examine the properties of 
the unitary quantum phase operator as well as the action-angle Wigner 
function.  
\end{abstract}
\newpage
\section{Introduction and Review}
The quantum mechanical operator realization of the classical phase 
observable,   
well-known as the historical quantum phase problem, is one of the oldest 
problems in quantum mechanics. In the quest for a correspondence between the 
classical action-angle variables and their quantum counterparts, Born, 
Heisenberg and Jordan have investigated\cite{1} the problem in the earliest   
days of quantum mechanics in the general  
perspective of building a theory of quantum canonical transformations and 
their unitary representations. The search for a quantum phase operator 
within this canonical perspective has specifically 
begun in one of Dirac's early  
works\cite{2} in 1927 where the principal motivation was to extend the 
principle of correspondence to that between the classical  
action-angle (AA) variables and their quantum counterparts. The quantum 
phase problem was then followed by the works of 
Heitler\cite{3} and Louisell\cite{4} where it was examined in terms of  
the quantization of the electromagnetic field. The introduction 
of {\it trigonometric~Hermitian} phase operators by Susskind and 
Glogower\cite{5} created a trigonometric approach to the phase problem. 
At this point, a landmark was made by the introduction of the coherent 
state formalism by Glauber\cite{6} 
and, with the development of laser physics in the 1960s, the theoretical and 
experimental investigation of the properties of quantum phase became   
mainstream in quantum optics. On the other hand, contemporary to 
Glauber's work,  
Carruthers and Nieto in their seminal paper\cite{7}, wisely entitled as 
{\it Phase and angle variables in quantum mechanics}, revived the interest  
on the canonical approach advocated in the early days of quantum mechanics.   
Since our interest in the current work is within the canonical perspective,  
we will refer the interested reader to, for instance, some recent reviews on 
the quantum 
phase as seen from the perspective of quantum optics.\cite{8,9} In search for 
the quantum counterpart of the classical AA pair, the canonical perspective  
in the quantum phase problem was furthered mainly by the works of 
Rocca and Siruge\cite{10}, Boyer and Wolf\cite{11}, 
Moshinsky and Seligman\cite{12}, as well as Luis and Sanchez-Soto\cite{13} 
and more recently by Lewis et al.\cite{14} 
In an earlier work\cite{15}, we introduced a different canonical-algebraic 
approach to the quantum phase problem from those in Ref's\,[11-14] 
by starting from the generalized discrete 
 unitary-cyclic finite (D) dimensional representations of the quantum phase 
space distribution 
functions in terms of Schwinger's operator basis.\cite{16,17} There are two 
crucial 
properties of these representations from the quantum phase operator point of 
view. The first one is that Schwinger's operator basis supports discrete cyclic 
finite dimensional subalgebraic representations with non-negative norms in 
the D dimensional Hilbert space $H_{D}$. These cyclic and admissible  
representations are known to be crucial for the existence of the phase operator 
in an arbitrary but finite dimensional algebra.  
On the other hand, the second crucial property is connected with the fact  
that the complete set of elements of the discrete finite dimensional 
cyclic Schwinger operator basis are the generalized dual representations 
of the standard Wigner-Kirkwood (WK)\cite{18} ones of the quantum 
phase space.\cite{15,16,17} 
Moreover, these elements are the generators of the discrete area  
preserving diffeomorphism on the two-dimensional toroidal lattice 
$\mbox{\ee Z}_{D} \times \mbox{\ee Z}_{D}$ which are known to respect  
the Fairlie-Fletcher-Zachos (FFZ) sine algebra.\cite{19} 
As the dimension $D$ is extended to 
infinity, a limit to continuum can be realized where  
the connection with Arnold's 
infinitesimal area preserving diffeomorphism\cite{20} on the continuous 
2-torus is established. Hence, the representations of the quantum phase space 
in terms 
of Schwinger's unitary-canonical operator basis paves a direct route to the 
algebraic formulation of the quantum phase operator in connection with the 
linear quantum canonical transformations (LCT). By this argument we imply 
that the algebraic formulation of the quantum phase problem is connected, 
through the Wigner-Weyl-Moyal (WWM) 
correspondence, with the existence of a canonical formalism of    
the quantum action-angle (AA) operators in the quantum phase space (QPS). 
This correspondence, although it will be shown to be manifest 
for arbitrary but finite dimensions 
leading to the finite dimensional algebraic realizations of the AA Wigner  
function, yields the desired correspondence between the quantum and the 
classical AA formalisms {\it only} in the transition to the continuum limit. 

The main purpose of this article is to extend the canonical-algebraic approach  
to the quantum phase problem in Ref.\,[15] by formulating this correspondence 
explicitly in terms of the generators of the LCT.  
The quantum AA operators will be found in terms of the 
generators of the LCT and it will be shown that the 
angle operator unitary-canonical to the quantum action will be identified as 
the unitary quantum phase operator. 

Here we review some relevant parts of Ref.\,[15] for the completeness of the 
present work. Some additional material is also included in the appendix. 
The duality relations between the discrete generalized 
WK phase space operator basis $\Delta(\vec{n})$ and the 
Schwinger operator basis $\hat{S}_{\vec m}$ can be expressed as\cite{15}  
\begin{equation}
\hat{\Delta} ({\vec n})=\frac{1}{D^{3/2}}\,\sum_{\vec m}\,e^{-i \gamma_{0}\,
({\vec m} \times {\vec n})}\,\hat{S}_{\vec m}~,\qquad 
\hat{S}_{\vec m}=\frac{1}{\sqrt{D}}\,\sum_{\vec n}\,
e^{i\gamma_{0}\,({\vec m} \times {\vec n})}\,\hat{\Delta}({\vec n})~, 
\label{WK1}
\end{equation}
where ${\vec m}=(m_1,m_2)$, ${\vec n}=(n_1,n_2)$ are vectors in 
$\mbox{\ee Z}_{D} \times \mbox{\ee Z}_{D}$; 
${\vec m} \times {\vec n} \equiv (m_1 n_2-m_2 n_1)$, $\gamma_0=2\pi/D$, 
with $D$ describing the dimension of the cyclic representations.  
Here the Schwinger operator basis $\hat{S}_{\vec m}$ is 
defined in terms of a finite dimensional unitary cyclic  
operator pair $(\hat{\cal U}, \hat{\cal V})$ such that  
\begin{equation}
\hat{\cal U}^{m_1}\,\hat{\cal V}^{m_2}=e^{i\gamma_{0}m_1 m_2}\,
\hat{\cal V}^{m_2}\, \hat{\cal U}^{m_1}, \qquad \quad 
\hat{S}_{\vec m}=
e^{-i\gamma_{0} m_1 m_2/2}\,\hat{\cal U}^{m_1}\,\hat{\cal V}^{m_2}~.  
\label{WK2}
\end{equation}
The $D$ dimensional cyclic eigenspace 
$\{\vert v \rangle_{k}\}_{0 \le k \le (D-1)}$ and 
$\{\vert u \rangle_{k}\}_{0 \le k \le (D-1)}$ 
of the operators $\hat{\cal U}, \hat{\cal V}$ satisfy 
\begin{equation}
\begin{array}{rl}
\hat{\cal U}\,\vert v\rangle_{k}=&
e^{i\gamma_{0} k}\,\vert v\rangle_{k}~,\qquad ~\,
\vert v\rangle_{k+D}=\vert v\rangle_{k} \\
\hat{\cal V}\,\vert u\rangle_{k}=&
e^{-i\gamma_{0} k}\,\vert u\rangle_{k}~,\qquad
\vert u\rangle_{k+D}=\vert u\rangle_{k} \\
\hat{\cal U}\,\vert u\rangle_{k}=&
\vert u\rangle_{k+1} \\
\hat{\cal V}\,\vert v\rangle_{k}=&
\vert v \rangle_{k+1}
\end{array}
\label{WK2.1}
\end{equation}
and define a unitary Fourier duality as 
\begin{equation}
\{\vert v\rangle\}=\hat{\cal F}\,\{\vert u\rangle\}~,\qquad 
{\rm where} \qquad (\hat{\cal F})_{k,k^\prime}=\frac{1}{\sqrt{D}}\,
e^{-i\gamma_{0}\,k\,k^\prime}~, 
\qquad \hat{\cal F}^{\dagger}=\hat{\cal F}^{-1}
\label{WK2.2}
\end{equation}
where, the dual picture implies a Fourier automorphism on $\hat{\cal U}$ and 
$\hat{\cal V}$ in a sequence of transformations as
\begin{equation}
{\hat{\cal U} \choose \hat{\cal V}} 
\stackrel{\hat{\cal F}}{\longrightarrow}
{\hat{\cal V} \choose \hat{\cal U}^{-1}} 
\stackrel{\hat{\cal F}}{\longrightarrow}
{\hat{\cal U}^{-1} \choose \hat{\cal V}^{-1}} 
\stackrel{\hat{\cal F}}{\longrightarrow}
{\hat{\cal V}^{-1} \choose \hat{\cal U}} 
\stackrel{\hat{\cal F}}{\longrightarrow}
{\hat{\cal U} \choose \hat{\cal V}}~. 
\label{WK2.3}
\end{equation}
It can be shown that this Fourier operator duality between $\hat{\cal U}$ 
and $\hat{\cal V}$ implies 
\begin{equation}
\hat{\cal F}\,\hat{S}_{\vec m}\,\hat{\cal F}^{-1}=
\hat{S}_{R_{\pi/2}:{\vec m}}~, \qquad \hat{\cal F}^{4}=1~, \qquad {\rm and} 
\qquad R_{\pi/2}^{4}=1
\label{WK2.4}
\end{equation}
where $R_{\pi/2}:{\vec m}=(-m_2,m_1)$ corresponds to a $\pi/2$ rotation   
of the vector ${\vec m}$ in the discrete phase space. 

Eq's\,(\ref{WK2}) and (\ref{WK2.1}) 
imply for the properties of the $\hat{S}_{\vec m}$ basis   
\begin{equation}
\begin{array}{rll}
\hat{S}_{\vec m}^{\dagger}=&\hat{S}_{-{\vec m}} \\
Tr\Bigl\{\,\hat{S}_{\vec m}\,\Bigr\}=&D\,\delta_{{\vec m},{\vec 0}} \\
\hat{S}_{\vec m}\,\hat{S}_{\vec m^{\prime}}=&
e^{i\,\gamma_{0}\,{\vec m}\times{\vec m^{\prime}}/2}\,
\hat{S}_{{\vec m}+{\vec m^{\prime}}} \\
(\hat{S}_{\vec m}\,\hat{S}_{{\vec m}^{\prime}})\,
\hat{S}_{{\vec m}^{\prime \prime}}=&
\hat{S}_{\vec m}\,(\hat{S}_{{\vec m}^{\prime}}\,
\hat{S}_{{\vec m}^{\prime \prime}}) \qquad &{\rm (associativity)}\\
\hat{S}_{\vec 0}=&\mbox{\ee I} &{\rm (unit~element)}\\
\hat{S}_{\vec m} \hat{S}_{-\vec m}=&\mbox{\ee I} &{\rm (inverse)}~. 
\end{array}
\label{WK3}
\end{equation}
The generalized discrete Wigner function $W({\vec n})$ in the  
physical state $\vert \psi \rangle \in H_{D}$ is defined by\cite{15,17} 
\begin{equation}
W({\vec n})=\langle \psi \vert \hat{\Delta}({\vec n}) \vert \psi \rangle
\label{WK5}
\end{equation}
where Eq.\,(\ref{WK5}) complies with all fundamental 
conditions that a generalized quantum phase space distribution should satisfy.  
The normalization of Eq.\,(\ref{WK5}) is based on an appropriate 
summation of the WK operator basis in Eq.\,(\ref{WK1}) over the discrete 
phase space vector ${\vec n}$. It is possible 
to use different normalizations when both(or one of the) labels are(is) 
continuous on the two dimensional torus. In these particular cases 
the phase space representations are based on $\mbox{\ee R} 
\times \mbox{\ee R}$ or $\mbox{\ee Z} \times \mbox{\ee R}$ respectively.  
Different normalizations are necessary for different choices of the phase 
space variables in order to obtain the appropriate continuum limit for the 
Wigner function. For instance, the symmetric normalization is necessary 
when the discrete phase space labels approach to the continuous ones 
simultaneously 
(i.e. $\mbox{\ee Z}_{D} \times \mbox{\ee Z}_{D}~~\to ~~
\mbox{\ee R} \times \mbox{\ee R}$)  
such as in the case of canonical pair of coordinate and  
momentum $x,p$ leading to the continuous phase space distribution 
$W(x,p)$. 
The limit to continuous action-angle Wigner function $W(J,\theta)$ is 
recovered\cite{15} when   
one of the phase space labels is real and the other remains to be an integer   
in the limit $D \to \infty$; hence 
$\mbox{\ee Z}_{D} \times \mbox{\ee Z}_{D}~~\to ~~
\mbox{\ee Z} \times \mbox{\ee R}$. 
In section.II we will base our formulation on the symmetric  
normalization as given in Eq.\,(\ref{WK1}); whereas, in section 4,  
the AA Wigner function will be examined using the appropriate 
asymmetric normalization with $\mbox{\ee Z} \times \mbox{\ee R}$ without any 
loss of generality. 

In section II we start the formulation of the LCT.  
The section II.A is devoted to the discrete scenario where the elements of 
LCT are in 
$SL(2,\mbox{\ee Z}_{D})$. Their action on Schwinger's discrete cyclic operator
basis is defined. The conditions of existence of a {\it unitary-canonical} 
partner to the generator of LCT are found for an arbitrary Hilbert 
space dimension and, provided such conditions are met, 
the existence of the unitary-canonical partner for each irreducible 
representation is demonstrated in the strong operator sense. 

In section II.B the continuous scenario is examined. The elements of the 
continuous LCT are examined within the context of the irreducible 
representations of $SL(2,\mbox{\ee R})$. The operators corresponding to the  
unitary canonical partners of continuous LCT  
within each one parameter subgroup of $SL(2,\mbox{\ee R})$ as well 
as the entire group are derived by the matrix elements of the diagonal 
representations of the corresponding group elements. The section III is 
devoted to the Hamiltonian system with its dynamical symmetry group 
corresponding to the group of LCT. The unitary canonical partner to the 
generators of LCT 
is identified as the unitary phase operator and their equations of motion  
are derived separately for Hamiltonians with continuous as well as discrete 
spectrum. 
The connections between the quantum AA formalism and the dynamical 
symmetry is established at the {\it operator~level}. The section IV is devoted 
to the construction of the AA-Wigner function. The continuous 
scenario is treated in section IV.a and the AA-Wigner function 
of the generalized oscillator with a discrete cyclic spectrum is presented 
in IV.b. The limit to the quantum harmonic oscillator AA formalism 
is also established. 
\section{Generators of the linear canonical transformations} 

\subsection{On the discrete toroidal lattice 
$\mbox{\ee Z}_{D} \times \mbox{\ee Z}_{D}$}
The unitary Fourier automorphism in Eq.\,(\ref{WK2.3}) implies the simplest 
discrete canonical transformation ${\vec m} \to R_{\pi/2}: {\vec m}$ on 
the phase space labels as given by Eq.\,(\ref{WK2.4}). It was shown 
in Ref.\,[15] that Eq's\,(\ref{WK2.3}) are a special case of a more 
general automorphic sequence produced by a unitary canonical transformation 
generator $\hat{\cal G}$ with $\hat{\cal G}^{\dagger}=\hat{\cal G}^{-1}$
where  
\begin{equation}
\begin{array}{rl}
\hat{\cal G}^{\dagger}\,\hat{U}\,\hat{\cal G}=\hat{S}_{\vec s}~,\qquad \qquad
\hat{\cal G}^{\dagger}\,\hat{V}\,\hat{\cal G}=\hat{S}_{\vec t} \\
\hat{\cal U} \, \stackrel{\hat{\cal G}}{\longrightarrow} \, \hat{S}_{\vec s}
\, \stackrel{\hat{\cal G}}{\longrightarrow} \,
\hat{S}_{s_1{\vec s}+s_2{\vec t}} \stackrel{\hat{\cal G}}{\longrightarrow}
...\\
\hat{\cal V} \, \stackrel{\hat{\cal G}}{\longrightarrow} \, \hat{S}_{\vec t}
\, \stackrel{\hat{\cal G}}{\longrightarrow} \,
\hat{S}_{t_1{\vec s}+t_2{\vec t}} \stackrel{\hat{\cal G}}{\longrightarrow}
...
\end{array}
\label{CT1}
\end{equation}
Such a unitary generator satisfies
\begin{equation}
\hat{\cal G}^{\dagger}\,\hat{S}_{\vec m}\,\hat{\cal G}=\hat{S}_{R:{\vec m}} 
~, \qquad {\rm where} \qquad R:{\vec m}={\vec m}^{\prime}=
(s_1 m_1+t_1 m_2, s_2 m_1+t_2 m_2)~
\label{CT2}
\end{equation} 
with $det R={\vec s} \times {\vec t}=1$ where ${\vec s}=(s_1,s_2)$ and 
${\vec t}=(t_1,t_2)$ are two arbitrary labeling vectors in 
$\mbox{\ee Z}_{D} \times \mbox{\ee Z}_{D}$. 
Hence $R \in SL(2,\mbox{\ee Z}_{D})$. 
Eq's\,(\ref{WK2.3}) and (\ref{WK2.4}) correspond to a special realization 
of Eq.\,(\ref{CT2}) when ${\vec s}=(0,1)$ and ${\vec t}=(-1,0)$. The 
application of $\hat{\cal G}$ leaves Eq's\,(\ref{WK3}) covariant. 

Using Eq.\,(\ref{CT2}) in Eq.\,(\ref{WK1}) it can be shown that 
$\hat{\cal G}$ generates discrete canonical transformations in the WK basis as  
\begin{equation}
\Delta({\vec n}^{\prime})=\hat{\cal G}^{\dagger}\,\Delta({\vec n})\,
\hat{\cal G}=
\Delta(R^{-1}:{\vec n})
\label{CT3}
\end{equation}
where $R^{-1}:{\vec n}={\vec n}^{\prime}=(t_2 n_1-t_1n_2,-s_2 n_1+s_1 n_2)$. 

The explicit form of $\hat{\cal G}$ and its irreducible representations 
have been studied in detail for the specific case of $D$ being a prime of the  
type $D=4k \pm 1$ where $k \in \mbox{\ee Z}$, in connection with the Schwinger 
operator basis by regarding $\hat{\cal G}$ as   
the generator of the time evolution (Hamiltonian) of the $SL(2,Z_{D})$ 
oscillator.\cite{21} Specifically,  
$\hat{\cal G}$ has cyclic generators which can be chosen as  
\begin{equation}
g_1=\pmatrix{1 & 1 \cr 0 & 1\cr}~,\qquad 
g_2=\pmatrix{g_0 & 0 \cr 0 & g_0^{-1}\cr}~,\qquad 
g_3=\Biggl\{\pmatrix{a & -b \cr b & a\cr}~, a^2+b^2=1 (mod D)\Biggr\}
\label{CT3j}
\end{equation}
with periods $D$, $D-1$ and $4k$ respectively. Here $g_0$ is a primitive 
element of $\mbox{\ee Z}_D$ where $g_0^{D-1}=1 (mod D)$.  

For such $D$, the explicit form of $\hat{\cal G}$ satisfying Eq's\,(\ref{CT1}),
or more compactly Eq.\,(\ref{CT2}), is given by\cite{21}   
\begin{equation}  
\hat{\cal G}=\hat{\cal G}(R)= 
\cases{\frac{\sigma(1)\sigma(\delta)}{D}\,\sum_{\vec m}\,
e^{i\gamma_0[t_1\,m_1^2+(t_2-s_1)m_1 m_2-s_2 m_2^2]/2\delta}\,\{m_1,m_2\}, 
&if $\delta \ne 0$, \cr 
\frac{\sigma(-2t_1)}{\sqrt{D}}\,\sum_{m_1}\,e^{i\gamma_0 m_1^2/2t_1}\,
\{m_1(s_1-1)/t_1, m_1\}, &if $\delta=0, t_1 \ne 0$, \cr 
\frac{\sigma(-2s_2)}{\sqrt{D}}\,\sum_{m_1}\,e^{-i\gamma_0\,m_1^2/2s_2}\,
\{m_1,0\}, &if $\delta=t_1=0, s_2 \ne 0$ \cr}
\label{CT3b}
\end{equation}
where it is defined that $\delta \equiv 2-s_1-t_2$ and 
$\{m_1,m_2\} \equiv \hat{S}_{\vec m}$. Here $\sigma(m)$ is the Gauss sum 
\begin{equation}
\sigma(m) \equiv \frac{1}{\sqrt D}\,\sum_{n=0}^{D-1}\,e^{i\gamma_0\,m\,n^2}
\label{CT3c}
\end{equation}
It can be seen by direct inspection that $\hat{\cal G}(R^{-1})=
\hat{\cal G}^{-1}(R)=\hat{\cal G}^{\dagger}(R)$ namely $\hat{\cal G}$ is 
unitary. 

Our main purpose in this section is to search for the condition of existence  
of a {\it unitary~canonical} partner $\hat{\cal O}$ 
to $\hat{\cal G}$ such that
\begin{equation}
\hat{\cal G}\,\hat{\cal O}=\Omega\,\hat{\cal O}\,\hat{\cal G}~,\qquad 
\vert \Omega \vert=1~,~~~~[\hat{\cal G},\Omega]=[\hat{\cal O},\Omega]=0~. 
\label{CT3d}
\end{equation}
If Eq.\,(\ref{CT3d}) is satisfied for some pure phase factor $\Omega$ and a 
unitary $\hat{\cal O}$, then we consider 
Eq.\,(\ref{CT3d}) as a generalized canonical  
commutation relation for the pair $\hat{\cal G}, \hat{\cal O}$. The 
Eq.\,(\ref{CT3d}) then implies that $\hat{\cal O}$ rotates the eigenspectrum  
of $\hat{\cal G}$ in a cyclic order and visa versa.  
It is beyond the scope of the manuscript to examine the characterization 
of the irreducible representations of the most general group defined by 
$\hat{\cal G}, \hat{\cal O}, \Omega$ in Eq.\,(\ref{CT3d}) above. Here, we will 
confine our attention to those relatively simpler cases leading to the   
unique irreducible representations of the group in which the operator 
$\hat{\cal O}$ becomes the unitary-canonical partner of the generalized  
canonical transformation generator $\hat{\cal G}$. For this purpose,   
let us start with the simplest case such that for some non-zero integers 
$a,b,c$ we have 
\begin{equation}
\hat{\cal G}^{a}=\hat{\cal O}^{b}=\Omega^{c}=1~. 
\label{CT3d1}
\end{equation}
Then we call the group defined by the elements 
$\hat{\cal G}^{\ell_1}\,\hat{\cal O}^{\ell_2}\,
\Omega^{\ell_3}$, where $\ell_1, \ell_2,\ell_3$ are integers defined 
$(mod\, a), (mod\, b), (mod\, c)$ respectively, as a discrete Heisenberg-Weyl 
group 
$\Gamma(a,b,c)$. An explicit calculation yields that, the group elements are 
uniquely defined only when $c$ divides both $a$ and $b$ (i.e. $a=c a^{\prime}, 
b=c b^{\prime}$ where $a^{\prime}, b^{\prime} \in \mbox{\ee Z}$). 
Furthermore, from Eq.\,(\ref{CT3d1}) and (\ref{CT3d}) we 
also have $\Omega^{a}=\Omega^{b}=\Omega^{c}=1$ which implies that 
$a,b,c$ must have 
a greatest common divisor $d$. These results imply that  
$a=d c^{\prime \prime} a^{\prime}, b=d c^{\prime \prime} b^{\prime}, 
c=d c^{\prime \prime}$ where $c^{\prime \prime} \in \mbox{\ee Z}$. 
Without loss of generality we will assume that $c^{\prime \prime}=1$. 
The group defined by the Schwinger operator basis $S_{\vec m}$ in 
Eq's\,(\ref{WK2}) and (\ref{WK2.1}) with general 
elements as $\hat{\cal U}^{m_1}\,\hat{\cal V}^{m_2}\,\omega^{m_3}$ 
where $m_1, m_2, m_3$ are integers $(mod D)$ is then a specific example of 
$\Gamma(a,b,c)$ with $a=b=c=D$. The number of irreducible representations 
of $\Gamma(a,b,c)$ with $a=d a^{\prime}, b=d b^{\prime}, c=d$ depends on 
the numbers $a^{\prime}, b^{\prime}$. For $a^{\prime}= b^{\prime}=1$ there is 
only one irreducible representation which is d-dimensional and is given by the 
Weyl matrices\cite{22}, 
\begin{equation}
\hat{\cal G} \quad \to \quad 
diag \,(1,\Omega,\Omega^2, \dots, \Omega^{d-1}) \qquad , \qquad  
\hat{\cal O} \quad \to \quad 
\pmatrix{0 & 0 & \dots & 0 & 1 \cr 
1 & 0 & \dots & 0 & 0 \cr 
0 & 1& 0 & \dots & 0 \cr 
\vdots & \vdots &\vdots & \ddots & 0\cr 
0 & 0 & \dots & 1& 0\cr}
\label{CT3i}
\end{equation}
where a unitary $\hat{\cal O}$ satisfying Eq.\,(\ref{CT3d}) 
exists as given in Eq.\,(\ref{CT3i}). For $a^{\prime}, b^{\prime} \ne 1$,  
the number of such irreducible representations is given by the product 
$a^{\prime} b^{\prime}$ and they are all d-dimensional. Up to a unitary 
equivalence, each irreducible representation is isomorphic to 
that in Eq.\,(\ref{CT3i}). The connection of direct product representation of 
$\Gamma(d,d,d)$ and its connection with the Chinese remainder theorem 
for the unique prime factorization of $d$ was studied recently 
in Ref.\,[23]. Each prime factor represents an independent physical 
degree of freedom 
allowing the extension of the phase space formalism presented here to more 
than one degrees of freedom.\cite{16,17} 
A similar decomposition can also be done for the more general 
group $\Gamma(a,b,d)$ with $a,b,d$ as defined above. The correspondence  
between the discrete QPS action-angle formalism with one degree of freedom and  
the classical one can be extended to more than one degrees of freedom at 
an algebraic level. Within the purpose of this article we will establish  
this correspondence only for the case with one degree of freedom and examine 
the larger degrees of freedom in a separate work.  

There is already an extensive literature on the representations of the 
discrete canonical transformations induced by 
$\hat{\cal G}=SL(2,\mbox{\ee Z}_{D})$. One particularly important limit 
in the discrete scenario is when $\hat{\cal G}$ is represented only by the 
rotational generator $g_3$ in Eq.\,(\ref{CT3j}) corresponding to the 
discrete fractional Fourier operator $\hat{\cal F}^{1/k}=\hat{\cal G}$  
such that $\hat{\cal G}^{4k}=\mbox{\ee I}$. This limit has been   
examined in detail both from the formal quantum mechanical\cite{24} and  
more applied, non-algebraic perspectives\cite{25,26}. For 
an arbitrary Hilbert space dimension  
the multiplicities of the four distinct 
eigenvalues of the fractional Fourier operator $\hat{\cal F}^{1/k}$ are not 
identical\cite{25} and neither $\hat{\cal F}$ nor $\hat{\cal F}^{1/4k}$ has 
exact unitary-canonical partner  
in the sense of $\hat{\cal O}$ satisfying Eq.\,(\ref{CT3d}). 

\subsection{Generators of the infinitesimal canonical transformations in 
$\mbox{\ee R} \times \mbox{\ee R}$} 
The modular group 
$SL(2,\mbox{\ee Z}_{D})$ does not have a proper continuous limit into  
$SL(2,\mbox{\ee R})$; hence, we cannot take the formal limit $D \to \infty$  
in Eq.\,(\ref{CT2}) to examine the continuous scenario. We will base  
the continuous representation of linear canonical transformations in 
$SL(2,\mbox{\ee R})$ in the formal sense on    
\begin{equation}
\hat{\cal G}_{\infty}^{\dagger}\,\hat{S}_{\vec \alpha}\,
\hat{\cal G}_{\infty}=
\hat{S}_{{\vec \alpha}^{\prime}}
\label{CT8}
\end{equation}
where ${\vec \alpha}=(\alpha_1,\alpha_2) \in \mbox{\ee R} \times 
\mbox{\ee R}$ is a continuous phase space vector, $\hat{S}_{\vec \alpha}$ 
are elements of the continuous Schwinger operator basis\cite{16} and 
${\vec \alpha}^{\prime}=R:{\vec \alpha}$ with $R \in SL(2,\mbox{\ee R})$ 
indicate the transformation matrix with real elements. From here on we  
will be confined to the continuous scenario  
in which we can drop the subscript $\infty$ from the 
canonical transformation generators $\hat{\cal G}_{\infty}$. 
The three  
one-parameter subgroups $g_j~,~~(j=1,2,3)$ of $sl(2,\mbox{\ee R})$, as 
conventionally 
represented by the three $2\times 2$ matrices, correspond to 
\begin{equation}
\Omega_{1}(\psi)=\pmatrix{cosh{\psi/2} & sinh{\psi/2} \cr
\sinh{\psi/2}& \cosh{\psi/2}\cr}~,\quad
\Omega_{2}(\theta)=\pmatrix{cos{\theta/2} & sin{\theta/2} \cr
-\sin{\theta/2}& \cos{\theta/2}\cr}~,\quad
\Omega_{3}(\varphi)=\pmatrix{e^{\varphi/2} & 0 \cr 
0& e^{-\varphi/2}\cr}~
\label{CT9}
\end{equation}
where $\Omega_{j} \in g_{j}$, $-\infty < \psi <\infty$, 
$-\pi < \theta <\pi$, and 
$-\pi \le \varphi \le \pi$. 
A generic group element $g \in SL(2,\mbox{\ee R})$ can be 
parameterized as 
\begin{equation}
g=\pmatrix{\alpha & \beta \cr \gamma & \delta}~, \qquad {\rm where} 
\qquad \det g=1 
\label{CT10}
\end{equation}
with $\alpha,\beta,\gamma,\delta \in \mbox{\ee R}$ being functions of 
$\psi, \theta, \varphi$.  

The three Hermitian operators $\hat{K}_{j}~,~(j=1,2,3)$ 
corresponding to the infinitesimal generators of the transformation   
in each subgroup respect the commutation relations 
\begin{equation}
[\hat{K}_{1},\hat{K}_{2}]=i\,\hat{K}_{3}~, \qquad
[\hat{K}_{2},\hat{K}_{3}]=i\,\hat{K}_{1}~, \qquad
[\hat{K}_{1},\hat{K}_{3}]=i\,\hat{K}_{2}~. 
\label{CT12}
\end{equation}
We are particularly interested in the continuous irreducible  
representation of $\hat{K}_{j}~,(j=1,2,3)$ in the canonical 
phase space 
parameterized by ${\vec \alpha}$ in Eq.\,(\ref{CT8}).
This particular representation of the generators is given by  
\begin{equation}  
\begin{array}{rl}
\hat{K}_{1}=&-i\,(\alpha_{1} \partial_{\alpha_{2}}+\alpha_{2}
\partial_{\alpha_{1}})/2 \\
\hat{K}_{2}=&-i\,(\alpha_{1} \partial_{\alpha_{2}}-\alpha_{2} 
\partial_{\alpha_{1}})/2 \\
\hat{K}_{3}=&-i\,(\alpha_{1}\partial_{\alpha_{1}}-
\alpha_{2}\partial_{\alpha_{2}})/2 \\
\end{array}
\label{CT13}
\end{equation}
where each irreducible representation acts on the 
Hilbert space of homogeneous polynomials 
of degree $2\ell$, and, a definite parity 
$\epsilon$ where $2\ell \in \mbox{\ee Z}$ with $\epsilon=\pm$ 
describing the odd(--) and even(+) parity. Hence we characterize those 
irreducible representations using the standard notation by 
$T_{\chi}(g_{j})$ where $\chi=(\ell,\epsilon)$ and the sector of the Hilbert 
space they belong to by $H_{\chi}$. 
Also, in the family $T_{\chi}(g_{j})$ we are particularly interested in  
the diagonal representations 
of each generator in Eq.\,(\ref{CT13}).  

\subsubsection{Diagonal representations of the $j$'th 
subgroup $g_j$} 
We will describe the eigenvectors $\vert e^{\chi}_{j}(\gamma_j)\rangle$ 
in the diagonal representations 
$T_{\chi}(g_j)$ characterized by a particular 
$j$ where $j=(1,2,3)$ with 
their corresponding projections on the canonical phase space  
$\langle {\vec \alpha} \vert e^{\chi}_{j}(\gamma_j)\rangle \equiv 
e_{j}^{\chi}({\vec \alpha},\gamma_j)$. Considering the simplest case of 
$\ell=0$ first, $\langle {\vec \alpha}\vert e^{\chi}_{j}(\gamma_j)\rangle$ 
are given by 
\begin{equation}
\begin{array}{rl}
T_{\chi}(\Omega_{1}): \qquad e_{1}^{\chi}({\vec \alpha},\gamma_1)=& 
C_{1}\,\Bigl(\frac{\alpha_{1}+\alpha_{2}}
{\alpha_{1}-\alpha_{2}}\Bigr)^{i\gamma_1}~, \\
T_{\chi}(\Omega_{2}): \qquad e_{2}^{\chi}({\vec \alpha},\gamma_2)=&
C_{2}\,\Bigl(\frac{\alpha_{1}+i\,\alpha_{2}}
{\alpha_{1}-i\alpha_{2}}\Bigr)^{\gamma_2}~, \\
T_{\chi}(\Omega_{3}): \qquad e_{3}^{\chi}({\vec \alpha},\gamma_3)=&
C_{3}\,\Bigl(\frac{\alpha_{1}}{\alpha_{2}})^{i\gamma_3}
\end{array}
\label{CT14}
\end{equation}
where $\gamma_j \in \mbox{\ee R}$ and $C_{j}$'s are constants based on an 
appropriate normalization by the inner product 
$<e_{j}^{\chi}(\gamma_j) \vert 
e_{j}^{\chi}(\gamma_j^{\prime})>=\delta(\gamma_{j}-\gamma^{\prime}_{j})$. 
Eq's\,(\ref{CT14}) imply that   
\begin{equation}
\hat{K}_{j}\,\vert e_{j}^{\chi}(\gamma_{j})\rangle =
\gamma_{j}\,\vert e_{j}^{\chi}(\gamma_{j})\rangle~.  
\label{CT15}
\end{equation}
Within each subgroup $g_{j}~,~~(j=1,2,3)$ we now define the  
unitary subgroup elements $\hat{\cal G}_{j} \in g_{j}$ such that 
\begin{equation}
\hat{\cal G}^{\Gamma_j}_{j}=e^{-i\,\Gamma_{j}\,\hat{K}_{j}}~,  
\label{CT16}
\end{equation}
where $\Gamma_j \in \mbox{\ee R}~,~~(j=1,2,3)$. The 
representations of Eq's\,(\ref{CT16}) in terms of $2\times 2$ matrices in the 
phase space ${\vec \alpha}={\alpha_1 \choose \alpha_2}$ are given by the 
$\Omega_{j}$'s in Eq.\,(\ref{CT9}), namely, the action of the each group 
element $\hat{\cal G}_{j}$ in Eq.\,(\ref{CT8}) is given by 
\begin{equation}
(\hat{\cal G}^{\Gamma_j}_{j})^{\dagger}\,\hat{S}_{\vec \alpha}\,
\hat{\cal G}^{\Gamma_j}_{j} \equiv \hat{S}_{\alpha^{\prime}}=
\hat{S}_{\Omega_{j}:\vec \alpha}~. 
\end{equation}
Within each subgroup $g_{j}$ there exists, in the Schwinger sense\cite{15}, 
a {\it special canonical} 
partner denoted by $\hat{\cal O}_{j}$ of 
$\hat{\cal G}_{j}$ such that  
\begin{equation}
\hat{\cal G}^{\Gamma_j}_{j}\,\hat{\cal O}^{\zeta_j}_{j}=
e^{-i\Gamma_j\,\zeta_j}\,
\hat{\cal O}^{\zeta_j}_{j}\,\hat{\cal G}^{\Gamma_j}_{j}~.
\label{CT18}
\end{equation}
Eq.\,(\ref{CT18}) implies that 
\begin{equation}
\hat{\cal G}^{\Gamma_j}_{j}\,\vert e^{\chi}_{j}(\gamma_j)\rangle=
e^{-i\Gamma_j\,\gamma_j}\,
\vert e^{\chi}_{j}(\gamma_j)\rangle~,\qquad
\hat{\cal O}^{\zeta_j}_{j}\,\vert e^{\chi}_{j}(\gamma_j)\rangle \sim  
\vert e^{\chi}_{j}(\gamma_j+\zeta_j) \rangle
\label{CT19}
\end{equation}
where $\sim$ sign in the equation on the right indicates that the equality 
holds up to an indeterminable phase factor which we consider to be  
irrelevant. Then Eq.\,(\ref{CT19}) provides a representation for the operator 
$\hat{\cal O}_{j}$ in 
the diagonal representation of $\hat{\cal G}_{j}$. 
To complete the picture, we consider for each subgroup $g_{j}$,  
the diagonal representation of the $\hat{\cal O}_{j}$ operator such that 
\begin{equation}
\hat{\cal O}^{\zeta_j}_{j}\,\vert f^{\chi}_{j}(\eta_j) \rangle=
e^{i\zeta_j \eta_j}\,\vert f^{\chi}_{j}(\eta_j)\rangle~. 
\label{CT20}
\end{equation}
From Eq's\,(\ref{CT18}) the action of $\hat{\cal G}^{\Gamma_j}_{j}$'s 
on this basis can be found as 
\begin{equation}
\hat{\cal G}^{\Gamma_j}_{j}\,\vert f^{\chi}_{j}(\eta_j)\rangle \sim 
\vert f^{\chi}_{j}(\eta_j+\Gamma_j) \rangle~.  
\label{CT21}
\end{equation}
The $\sim$ indicates again that there is an overall indeterminable 
phase which we ignore without any loss of generality. Then  
Eq's\,(\ref{CT18}-\ref{CT21}) completely determine, in the weak  
sense, the properties 
of the operator pair $(\hat{\cal G}, \hat{\cal O})$ within the Hilbert space 
of each subgroup $g_j$ as three different realizations of the Schwinger 
operator basis. The connection between the eigenbasis 
$\vert e^{\chi}_{j}(\gamma_j)\rangle$ and 
$\vert f^{\chi}_{j}(\eta_j)\rangle $ 
is given by the Fourier transformation 
\begin{equation}
\vert f^{\chi}_{j}(\eta_j)\rangle =\int\,d\gamma_j\,e^{-i\,\gamma_j\,\eta_j}\,
\vert e^{\chi}_{j}(\gamma_j)\rangle~,\qquad 
\vert e^{\chi}_{j}(\gamma_j)\rangle=\int\,d\eta_j\,e^{i\eta_j\,\gamma_j}\,
\vert f^{\chi}_{j}(\eta_j)\rangle
\label{CT22}
\end{equation}
where in short notation $\vert f_{j}^{\chi}(\eta_j)\rangle=\hat{\cal F}\,
\vert e_{j}^{\chi}(\gamma_j)\rangle$ with 
$(\hat{\cal F})_{\eta_j,\gamma_j}=\Bigl(\langle f_{j}^{\chi}(\eta_j)\vert 
e_{j}^{\chi}(\gamma_j)\rangle\Bigr)$  
describing the matrix elements parameterized by $(\eta_j,\gamma_j)$ 
of the unitary Fourier operator $\hat{\cal F}$. Via Eq.\,(\ref{CT22}) a 
Fourier automorphism is implied between the two eigenspaces for each $j$ as   
\begin{equation}
\vert e^{\chi}_{j}(\gamma_j)\rangle\stackrel{\hat{\cal F}}{\longrightarrow} 
\vert f^{\chi}_{j}(\eta_j)\rangle \stackrel{\hat{\cal F}}{\longrightarrow} 
\vert e^{\chi}_{j}(-\gamma_j)\rangle \stackrel{\hat{\cal F}}{\longrightarrow} 
\vert f^{\chi}_{j}(-\eta_j)\rangle \stackrel{\hat{\cal F}}{\longrightarrow} 
\vert e^{\chi}_{j}(\gamma_j)\rangle~. 
\label{CT23}
\end{equation}

\subsubsection{Diagonal representations of the entire group $g$} 

We now shift our attention from the parameterization of the diagonal  
representations  
to those of the 
entire group $g$ of which the three-parameter group element will  
be denoted by $\hat{\cal G}$. 
For convenience of the calculations we adopt the unitary canonical  
form for $\hat{\cal G}$ as\cite{27}
\begin{equation}
\hat{\cal G}^{\Lambda}=e^{-i\,{\vec \Lambda}.\vec{\hat{K}}}~,\qquad 
{\vec \Lambda}=(\Lambda_1,\Lambda_2,\Lambda_3)~,\qquad
\vec{\hat{K}}=(\hat{K}_1, \hat{K}_2, \hat{K}_3)
\label{CT24}
\end{equation}
where ${\vec \Lambda}$ is defined on the $sl(2,\mbox{\ee R})$ invariant group 
manifold characterized by the invariant 
$\Lambda^{2}=\Lambda_1^2+\Lambda_3^2-\Lambda_2^2$. Adopting the particular 
parameterization $\Lambda_1=\Lambda\,
\sin{a} \cosh{b}$, $\Lambda_2=\Lambda\,\sin{a} \sinh{b}$, 
and $\Lambda_3=\Lambda\,\cos{a}$, Eq.\,(\ref{CT24}) can be obtained, 
let's say, from 
$\hat{\cal G}_3^{\Lambda}$ in Eq.\,(\ref{CT16}) by the 
unitary transformation
\begin{equation}
\hat{\cal G}^{\Lambda}=(\hat{\cal T}_{23}^{(-a,b)})^{\dagger}\,
\hat{\cal G}_3^{\Lambda}\,\hat{\cal T}_{23}^{(-a,b)}
~,\qquad {\rm where} \qquad 
\hat{\cal T}_{23}^{(-a,b)}=\hat{\cal G}^{-a}_{2}\,\hat{\cal G}^{b}_{3}~. 
\label{CT25}
\end{equation} 
Since $\hat{\cal G}^{\Gamma}$ in 
Eq.\,(\ref{CT24}) is an element of $sl(2,\mbox{\ee R})$, the 
irreducible representations also act on the homogeneous 
polynomials of degree $2\ell$ 
and parity $\epsilon=\pm$ in $H_{\chi}$. 

i) Continuous diagonal representations:

Similar to Eq's\,(\ref{CT15}) and to the first set in Eq's\,(\ref{CT19}) 
for the subgroups, we now seek the 
eigenvectors $\vert h^{\chi}(\gamma)\rangle \in H_{\chi}$ 
of $\hat{\cal G}^{\Lambda}$ such that 
\begin{equation}
\hat{\cal G}^{\Lambda}\,\vert h^{\chi}(\gamma)\rangle= e^{i\,A}\,
\vert h^{\chi}(\gamma)\rangle ~, \qquad A \in \mbox{\ee R} 
\label{CT26}
\end{equation}
where $A$ and $\vert h^{\chi}(\gamma)\rangle$ are to be found from the 
eigenproblem in 
Eq.\,(\ref{CT26}). Eq.\,(\ref{CT25}) suggests that 
\begin{equation}
\vert h^{\chi}(\gamma) \rangle=
(\hat{\cal T}_{23}^{(-a,b)})^{\dagger}\,
\vert e_{3}^{\chi}(\gamma) \rangle
\label{CT27}
\end{equation}
where $A=-\Lambda$ in Eq.\,(\ref{CT26}). Hence $\vert h^{\chi}(\gamma) \rangle$ 
spans the eigenspace of the unitary operator in Eq.\,(\ref{CT24}) with 
$\Lambda, \gamma \in \mbox{\ee R}$. The orthonormality of the eigenbasis
$\vert h^{\chi}(\gamma) \rangle$ is guaranteed by the unitary transformation 
in Eq.\,(\ref{CT27}) and the orthonormality of the eigenbasis 
$\vert e_{3}^{\chi}(\gamma)\rangle$. A phase space representation for 
$\vert h^{\chi}(\gamma)\rangle$ similar to  
Eq's\,(\ref{CT14}) can be found by projecting it on the phase space 
vector ${\vec \alpha}$ as 
\begin{equation}
\langle {\vec \alpha}\vert h^{\chi}(\gamma)\rangle \equiv 
h^{\chi}({\vec \alpha}, \gamma)=\langle {\vec \alpha}\vert 
(\hat{\cal T}_{23}^{(-a,b)})^{\dagger}\,
\vert e_{3}^{\chi}(\gamma) \rangle~ \qquad 
\hat{\cal T}_{23}^{(-a,b)}{\alpha_1 \choose \alpha_2}=
\Omega_{3}(b)\,\Omega_{2}(a)\,{\alpha_1 \choose \alpha_2}
\label{CT28}
\end{equation}
where $\Omega_{2}(a)$ and $\Omega_{3}(b)$ are implied by Eq.\,(\ref{CT9}). 
The {\it unitary-canonical} partner of 
$\hat{\cal G}$ can be found similarly as it was done for the subgroups.
Defining $\hat{\cal O}$ such that 
\begin{equation}
\hat{\cal G}^{\Lambda}\,\hat{\cal O}^{\zeta}=e^{-i\Lambda\,\zeta}\,
\hat{\cal O}^{\zeta}\,\hat{\cal G}^{\Lambda}
\label{CT28.b}
\end{equation}
we find
\begin{equation}
\begin{array}{rl}
\hat{\cal G}^{\Lambda}\,\vert h^{\chi}(\gamma)\rangle=&
e^{-i\Lambda\,\gamma}
\,\vert h^{\chi}(\gamma)\rangle ~, \qquad 
\hat{\cal O}^{\zeta}\,\vert h^{\chi}(\gamma)\rangle \sim 
\vert h^{\chi}(\gamma+\zeta)\rangle \\
\hat{\cal O}^{\zeta}\,\vert k^{\chi}(\eta)\rangle=&e^{i\zeta \eta}\,
\vert k^{\chi}(\eta)\rangle~,\qquad 
\hat{\cal G}^{\Lambda}\,\vert k^{\chi}(\eta)\rangle \sim
\vert k^{\chi}(\eta+\Lambda)\rangle 
\end{array}
\label{CT29}
\end{equation}
where we will again neglect the overall phases in the second column of the 
relations above. The basis vectors in Eq.\,(\ref{CT29}) are connected by the 
Fourier transformation
\begin{equation}
\vert k^{\chi}(\eta)\rangle=\int\,d\gamma\,e^{-i\gamma \eta}\,
\vert h^{\chi}(\gamma)\rangle~,\qquad 
\vert h^{\chi}(\gamma)\rangle=\int\,d\eta\,e^{i\gamma \eta}\,
\vert k^{\chi}(\eta)\rangle
\label{CT30}
\end{equation}
and a similar automorphism to Eq.\,(\ref{CT23}) at the vector level between 
$\vert h^{\chi}(\gamma)\rangle$ and $\vert k^{\chi}(\eta)\rangle$ as well as 
to Eq.\,(\ref{WK2.3}) 
at the operator level between $\hat{\cal G}$ and $\hat{\cal O}$ can be 
written.  

ii) Discrete diagonal representations:

In examining the discrete representations of the entire group we 
start with the diagonal ones $\vert e_{2}^{\chi}(m)\rangle$ of $\hat{K}_2$  
and associate with them the eigenfunctions 
\begin{equation}
\langle {\vec \alpha}\vert e_{2}^{\chi}(m)\rangle \equiv 
e_{2}^{\chi}({\vec \alpha};m)=N_{2}^{(\ell)}\,(\alpha_1^2+\alpha_2^2)^\ell\,
(\frac{\alpha_1+i\alpha_2}{\alpha_1-i\alpha_2})^m
\label{DR1}
\end{equation}
where $N_2^{(\ell)}$ is a normalization based on an inner product 
$\langle e_{2}^{\chi}({\vec \alpha}; m)\vert e_{2}^{\chi}({\vec \alpha}; 
m^{\prime})\rangle=\delta_{m,m^\prime}$. 
For the state in Eq.\,(\ref{DR1}) we have
\begin{equation}
\begin{array}{rl}
\hat{K}_2\,\vert e_{2}^{\chi}(m)\rangle=&m\,
\vert e_{2}^{\chi}(m)\rangle \\ 
\hat{K}_+\,\vert e_{2}^{\chi}(m)\rangle=&-(\ell-m)\,
\vert e_{2}^{\chi}(m+1)\rangle \\
\hat{K}_-\,\vert e_{2}^{\chi}(m)\rangle=&(\ell+m)\,
\vert e_{2}^{\chi}(m-1)\rangle
\end{array}
\label{DR2}
\end{equation}
where $\hat{K}_{\pm}=(\hat{K}_1 \pm i\hat{K}_3)$ 
and the Casimir element $\hat{K}^2=1/2(\hat{K}_{+}\hat{K}_{-}+
\hat{K}_{-}\hat{K}_{+})-K_2^2$ has eigenvalue $\ell(\ell+1)$. In terms of 
the $\ell \ne 0$ representation in Eq.\,(\ref{DR1}) and (\ref{DR2}), the 
continuous representations that we used in Eq.\,(\ref{CT14}) correspond to the 
$\ell =0$ case. A general 
eigenvector for a group element in Eq.\,(\ref{CT24}) can be found 
similarly as in the continuous case by redefining 
${\vec \Lambda}=(\Lambda_1,\Lambda_2,\Lambda_3)$ as, for instance, 
$\Lambda_1=\Lambda\,\sinh{a^\prime}\sin{b^\prime},~
\Lambda_2=\Lambda\,\cosh{a^\prime},~
\Lambda_3=\Lambda\,\sinh{a^\prime}\cos{b^\prime}$ and the transformation  
\begin{equation}
\hat{\cal G}^{\Lambda}=(\hat{\cal T}_{12}^{(a^\prime,b^\prime)})^{\dagger}
\,\hat{\cal G}_2^{\Lambda}\,\hat{\cal T}_{12}^{(a^\prime,b^\prime)}
~,\qquad {\rm where} \qquad 
\hat{\cal T}_{12}^{(a^\prime,b^\prime)}=
\hat{\cal G}^{a}_{1}\,\hat{\cal G}^{b}_{2}~. 
\label{DR3}
\end{equation}
The eigenvectors $\vert h^{\chi}(m)\rangle$ of a general group element 
$\hat{\cal G}^{\Lambda}$ are then associated with the functions 
in $H_{\chi}$ 
\begin{equation}
\langle {\vec \alpha}\vert h^{\chi}(m)\rangle \equiv 
h^{\chi}({\vec \alpha};m)=
\langle {\vec \alpha} \vert 
(\hat{\cal T}_{12}^{(a^\prime,b^\prime)})^{\dagger}\,
\vert e_{2}^{\chi}(m)\rangle~,\qquad   
\hat{\cal T}_{12}^{(a^\prime,b^\prime)} \, {\alpha_1 \choose \alpha_2}=
\Omega_{2}(-b^{\prime})\,\Omega_{1}(a^{\prime})\,{\alpha_1 \choose \alpha_2}
\label{DR4}
\end{equation}
where $\Omega_{2}(-b)$ and $\Omega_{1}(a)$ are implied by Eq's\,(\ref{CT9}). 
To establish the representation of the unitary canonical pair 
$(\hat{\cal G},\hat{\cal O})$ in the discrete case we consider 
\begin{equation}
\hat{\cal G}^{\Lambda}\,\hat{\cal O}^{r}=e^{-i\Lambda\,r}\,
\hat{\cal O}^{r}\,\hat{\cal G}^{\Lambda}
\label{DR5}
\end{equation}
corresponding to a change $\zeta \to r \in \mbox{\ee Z}$ in 
Eq.\,(\ref{CT28.b}). 
For the system of eigenvectors we find  
\begin{equation}
\begin{array}{rl}
\hat{\cal G}^{\Lambda}\,\vert h^{\chi}(m)\rangle=&
e^{-i\Lambda\,m}
\,\vert h^{\chi}(m)\rangle~, \qquad
\hat{\cal O}^{r}\,\vert h^{\chi}(m)\rangle \sim
\vert h^{\chi}(m+r)\rangle \\
\hat{\cal O}^{r}\,\vert k^{\chi}(\eta)\rangle=&e^{i\eta r}\,
\vert k^{\chi}(j,\eta)\rangle~,\qquad ~~~~
\hat{\cal G}^{\Lambda}\,\vert k^{\chi}(\eta)\rangle \sim
\vert k^{\chi}(\eta+\Lambda)\rangle
\end{array}
\label{DR6}
\end{equation}
where $\Lambda, \eta \in \mbox{\ee R}$. The eigenspaces are connected 
by the (discrete and continuous) Fourier transformation as 
\begin{equation}
k^{\chi}({\vec \alpha};\eta)=\sum_{m}e^{-i\eta\,m}\,
h^{\chi}({\vec \alpha};m)~,\qquad
h^{\chi}({\vec \alpha};m)=\int\,d\eta\,e^{i m \eta}\,
k^{\chi}({\vec \alpha};\eta)~. 
\label{DR7}
\end{equation}
The Eq's\,(\ref{DR6}) and (\ref{DR7}) conclude our brief treatment of the 
unitary continuous phase space  
representations of the canonical $\hat{\cal G}, \hat{\cal O}$ pair in the 
weak matrix element sense. 

\section{Implications for the Hamiltonian systems with dynamical  
symmetry group and generalized action-angle operators}
Recently an algebraic approach was studied by Wang and Chu\cite{28} in the 
solution of the one dimensional inverse problem for the Hamiltonian systems 
with dynamical group symmetry. In the simplest case of one dimension, the 
inverse problem reduces to that of finding 
an operator regarding the dynamical symmetry group of 
the system such that it will be invariant under either a certain subgroup 
or the entire group of canonical transformations. 
The group of canonical transformations then naturally reflects the  
properties of the dynamical 
symmetry of the system. For one dimensional {\it autonomous} systems, 
the invariant operator under canonical transformations corresponds to the  
generalized {\it action} operator, by which the Hamiltonian of the 
entire system can be fully described. In the following, we 
will examine the quantum canonical transformation group as the 
dynamical symmetry group for an Hamiltonian system and derive the 
equations of motion describing the time evolution of the generalized quantum 
action-angle operators. 

Let us assume that the Hamiltonian $\hat{\cal H}$ describing the 
dynamics of a system with one degree of freedom is represented by  
\begin{equation}
\hat{\cal H} \equiv {\cal H}(\hat{K}_{1},\hat{K}_{2},\hat{K}_{3})
\label{DSG1}
\end{equation}
where $\hat{K}_{j}$'s are the Hermitian generators of the  
$sl(2,\mbox{\ee R})$ canonical transformation algebra in Eq.\,(\ref{CT24}). 
The simplest but a sufficiently general example 
of an Hamiltonian with a dynamical group symmetry can then be obtained 
if the Hamiltonian is a real function of the linear superposition of 
$\hat{K}_{j}$'s as, 
\begin{equation}
\hat{\cal H}={\cal H}(P\,\hat{K}_1+Q\,\hat{K}_2+R\,\hat{K}_3)~,\qquad 
P,Q,R \in \mbox{\ee R}~. 
\label{DSG2}
\end{equation}
Here, we do not make any assumptions aside from the one that ${\cal H}$ is a 
real valued and a well-behaved function. Our focus will be on those   
unitary elements of the dynamical symmetry group that can be described by  
\begin{equation} 
\hat{\cal G}^{\Gamma}=
e^{-i\,\Gamma\,(P\,\hat{K}_1+Q\,\hat{K}_2+R\,\hat{K}_3)}
\label{DSG3}
\end{equation}
where the Hamiltonian in Eq.\,(\ref{DSG2}) becomes 
\begin{equation}
\hat{\cal H}={\cal H}(i\,\frac{\partial}{\partial \Gamma})\,
\hat{\cal G}^{\Gamma} 
\Bigr\vert_{\Gamma=0}~. 
\label{DSG4}
\end{equation}
The time dependent eigenvectors $\vert \psi^{\chi}_{\gamma} (t)\rangle$ 
of $\hat{\cal H}$ and the eigenenergies in 
the continuous representation are given by 
\begin{equation}
\langle {\vec \alpha} \vert \psi_{\gamma}^{\chi}(t) \rangle=
e^{i E_{\gamma} t}\,
h^{\chi}({\vec \alpha},\gamma) ~, \qquad E_{\gamma}={\cal H}(\gamma) 
\label{DSG5}
\end{equation}
where $h^{\chi}({\vec \alpha},\gamma)$ is given by Eq.\,(\ref{CT28}). 
Here, the unitary canonical partner $\hat{\cal O}$ to 
$\hat{\cal G}$ plays the role of the unitary angle operator and is defined 
by Eq.\,(\ref{CT28.b}). The time 
evolution of $\hat{\cal O}^{\zeta}$ is given by 
\begin{equation}
i\,\frac{d}{dt}\,\langle \psi_{\gamma^{\prime}}\vert 
\hat{\cal O}^{\zeta}\vert \psi_{\gamma}\rangle=
\langle \psi_{\gamma^{\prime}}\vert\,
[\hat{\cal H},\hat{\cal O}^{\zeta}]\,
\vert \psi_{\gamma^{\prime}}\rangle=-\langle \psi_{\gamma^{\prime}}\vert\,
\hat{\cal O}^{\zeta}\vert \psi_{\gamma}\rangle\,
\Bigl\{{\cal H}(\gamma+\zeta)-{\cal H}(\gamma)\Bigr\}~. 
\label{DSG6}
\end{equation}
If we symbolically associate an {\it Hermitian} phase operator   
at $\zeta=0$ with  
\begin{equation}
\hat{\phi}=-i\,\frac{d}{d\zeta}\,\hat{\cal O}^{\zeta}\,
\Bigr\vert_{\zeta=0} 
\label{DSG7}
\end{equation} 
Eq.\,(\ref{DSG6}) becomes, in terms of $\hat{\phi}$
\begin{equation}
\lim_{\zeta \to 0}\,\langle \psi_{\gamma-\zeta}\vert \frac{d}{dt}\,
\hat{\phi} \vert \psi_{\gamma} \rangle=-\frac{d E_{\gamma}}{d\gamma}\,
\lim_{\zeta \to 0}\,\langle \psi_{\gamma-\zeta}
\vert \psi_{\gamma}\rangle
\label{DSG8}
\end{equation}
which is the quantum analog of the classical equation of motion for the 
canonical angle variable. Hence the unitary operator $\hat{\cal O}$ 
should be considered as the {\it quantum~angle~operator} which is the unitary 
canonical partner of the action operator $\hat{\cal G}$. 

The properties of the dynamical group symmetry can also be examined  
by using the discrete representation. In this case the 
Eq.\,(\ref{DSG4}) is still valid whereas (\ref{DSG5}) becomes  
\begin{equation}
\langle {\vec \alpha} \vert \psi_{m}^{\chi}(t) \rangle=e^{i E_{m} t}\,
h^{\chi}({\vec \alpha};m) ~, \qquad E_{m}={\cal H}(m)
\label{DSG9}
\end{equation}
where $\langle {\vec \alpha} \vert \psi_{m}^{\chi}(t) \rangle$ are the time 
dependent eigenvectors, 
$h^{\chi}({\vec \alpha};m)$ is given by Eq.\,(\ref{DR4})  
and $E_{m}={\cal H}(m)$ is the discrete eigenenergy spectrum 
depending on the discrete eigenstate index $m$. The appropriate angle 
operator $\hat{\cal O}^{r}$ in the discrete case has been studied in 
Eq.\,(\ref{DR5}) 
and (\ref{DR6}). The nature of the 
operator $\hat{\cal O}^{r}$ becomes more transparent  
if we examine the derivative of Eq.\,(\ref{DR5}) 
with respect to $\Lambda$ at $\Lambda=0$. This can be readily evaluated as 
\begin{equation}
\frac{d}{d\Lambda} \to {\rm Eq}.\,(\ref{DR5})\Bigr\vert_{\Lambda=0}
 \Longrightarrow   [{\vec n}.\vec{\hat{K}},\hat{\cal O}^{r}]=-r\,
\hat{\cal O}^{r}
\label{DSG6.a}
\end{equation}
which is the generalized Susskind-Glogower-Carruthers-Nieto commutation 
relation\cite{5,7,15} for the generalized {\it radial~number~operator} 
${\vec n}.\vec{\hat{K}}$ and the generalized unitary phase operator 
$\hat{\cal O}$. The time evolution of 
$\hat{\cal O}^{r}$ is then
\begin{equation}
i\,\frac{d}{dt}\,\langle \psi_{m^\prime}^{\chi}\vert
\hat{\cal O}^{r}\vert \psi_{m}^{\chi}\rangle=
\langle \psi_{m^\prime}^{\chi}\vert\,
[\hat{\cal H},\hat{\cal O}^{r}]\,
\vert \psi_{m}^{\chi}\rangle=-\langle \psi_{m^\prime}^{\chi}\vert\,
\hat{\cal O}^{r}\vert \psi_{m}^{\chi}\rangle\,
\Bigl\{{\cal H}(m+r)-{\cal H}(m)\Bigr\}
\label{DSG6.b}
\end{equation}
which is the equation of motion for the unitary canonical angle operator 
of the generalized oscillator with a discrete spectrum $E_{m}={\cal H}(m)$. 
Here, two results are in order. The first is that we have found a 
correspondence between the classical and quantum AA formalisms for Hamiltonian 
systems with a dynamical group symmetry of the type given by 
Eq.\,(\ref{DSG2}) and (\ref{DSG3}). 
The second result is the equivalence 
of the quantum action operator to the generators of canonical transformations 
as well as that of the unitary-canonical angle operator to the  
unitary-canonical phase operator.
  
Although the continuous and discrete representations of the AA operators 
are similar, 
depending on the continuous/discrete nature of the eigenenergy spectrum, one 
or the other is more convenient in the formulation of a physical problem.   
This will be more transparent in the next section when we discuss the 
action-angle formalism of a generalized oscillator. 
\subsection{Implications for the generalized oscillator Hamiltonian 
and the quantum phase operator}
In this and the following sections we will refer to the appendix which 
includes some relevant parts of Ref.\,[15]. There, 
it was shown that the quantum harmonic oscillator (QHO) algebra 
is recovered 
in the infinite dimensional limit (i.e. $D \to \infty$ hence $q \to 1$) 
of the admissible q-oscillator algebraic realization in 
Eq's\,(A.4). The importance of the naturally emerging admissible 
q-oscillator realizations is that they admit an algebraic formulation of 
the quantum phase problem and also provide a natural basis to examine 
the harmonic oscillator phase in the infinite dimensional limit of the 
algebra in Eq's\,(A.4). Respecting the historical development,    
we will nevertheless start with a brief outline of the phase problem  
using the dynamical continuous symmetry group of the QHO.  
The generators 
$\hat{K}_i~,~~(i=1,2,3)$ 
of the dynamical $sl(2,\mbox{\ee R})$ symmetry of the QHO  
in the $\hat{x},\hat{p}$ representation are given 
by  
\begin{equation}
\hat{K}_1=\frac{1}{4}\,(\hat{x}^2-\hat{p}^2) ~, \qquad  
\hat{K}_2=\frac{1}{4}\,(\hat{x}^2+\hat{p}^2) ~, \qquad 
\hat{K}_3=\frac{1}{4}\,(\hat{x}\hat{p}+\hat{p}\hat{x}) ~, \qquad 
[\hat{x},\hat{p}]=i 
\label{OSC1}
\end{equation}
where the generators respect Eq's\,(\ref{CT12}).  
Our first attempt will be to find the unitary canonical partner to 
$\hat{K}_3$ in Eq's\,(\ref{OSC1}). With $\hat{p} \to -i \partial/\partial x$, 
the eigenproblem for $\hat{K}_3$ yields
\begin{equation}
\hat{K}_3\,\vert \psi_3(\gamma_3)\rangle=
\gamma_3\,\vert \psi_3(\gamma_3)\rangle~, \qquad 
\gamma_3 \in \mbox{\ee R}~, \qquad \langle x\vert \psi_3(\gamma_3)\rangle=N_3\,
x^{(2i\gamma_3-1/2)}
\label{OSC2}
\end{equation}
where $N_3$ is a normalization based on an inner product. Hence, for 
$\hat{\cal G}_3^{\Gamma_3}=e^{-i\Gamma_3\,\hat{K}_3}$, 
\begin{equation}
\hat{\cal G}_3^{\Gamma_3}\,\vert \psi_3(\gamma_3)\rangle=
e^{-i\Gamma_3\,\gamma_3}\,
\vert \psi_3(\gamma_3)\rangle~. 
\label{OSC3}
\end{equation}
The unitary canonical partner $\hat{\cal O}_{3}$ to $\hat{\cal G}_3$, such that 
Eq.\,(\ref{CT18}) is satisfied for $i=3$, can be found following the steps 
leading to Eq's\,(\ref{CT19}-\ref{CT21}). The $\hat{\cal O}_{3}^{\zeta}$ 
operator for an arbitrary and real $\zeta$ is given by 
\begin{equation}
\begin{array}{rl}
\hat{\cal O}^{\zeta}_{3}=&\int_{-\infty}^{\infty}\,d\gamma_3\, \vert 
\psi(\gamma_3+\zeta)\rangle\,\langle \psi(\gamma_3)\vert~,\qquad {\rm or~
equivalently~by} \\
\hat{\cal O}^{\zeta}_{3}=&
\int_{-\infty}^{\infty}\,d\eta_3\,e^{i\zeta\,\eta_3}\,
\vert{\tilde \psi}(\eta_3)\rangle\langle {\tilde \psi}(\eta_3)\vert~,\qquad 
{\rm where} \\ 
\vert{\tilde \psi}(\eta_3)\rangle=&
\int_{-\infty}^{\infty}\,d{\gamma_3}\,e^{-i\eta_3\gamma_3}\,
\vert \psi(\gamma_3)\rangle
\end{array}
\label{OSC4}
\end{equation}
namely, the existence of $\hat{\cal O}_{3}$ is manifested by the presence of 
a complete spectrum of $\hat{K}_3$ on the real axis and, in return, 
$\hat{\cal O}_{3}$ and $\hat{\cal G}_3$ are connected by a Fourier 
automorphism. 
A similar procedure can also be applied to $\hat{K}_1$ in Eq.\,(\ref{OSC1}) 
since the eigenspectrum of this operator also spans the symmetric positive 
and negative values on the entire real axis. However, there is a problem 
with the $\hat{K}_2$ operator. Because of the fact that $\hat{K}_2$ 
in Eq's\,(\ref{OSC1}) is a non-negative operator its eigenspectrum spans 
only the positive real axis. Hence, the Fourier automorphism is not 
applicable to $\hat{K}_2$ and in return, the unitary canonical partner to 
$\hat{\cal G}_2$ cannot be found. This  
problem has been attacked from a completely different perspective a 
long time ago in the elegant work of Boyer and Wolf\cite{11} where they  
made use of the unitary isomorphism between  
the {\it radial} representation of the dynamical $SL(2,\mbox{\ee R})$ symmetry 
group of the multi-dimensional QHO with an added centrifugal term of arbitrary  
strength and the representation of 
the same group on the unit circle.  
Through this unitary mapping the space of square integrable functions on the 
unit circle is an inner product space endowed with a translationally  
invariant non-local measure. However, the drawback of this elegant 
method is that, the unitary irreducible representations are not  
single-valued under full rotations on the unit circle and this applies 
particularly to the standard one-dimensional quantum harmonic oscillator.  

The phase problem in the QHO being the central theme of this work, we suggest 
here and in the following sections 
an alternative and, perhaps, a formally simpler way of looking into 
the phase problem in the QHO. We will now start with the admissible 
q-oscillator realization in Eq.\,(A.4) with the operators 
$\hat{A}, \hat{A}^{\dagger}$ and $\hat{Q}$ and approach the QHO algebra by  
extending $D$ to infinity. We formally express the 
generalized oscillator Hamiltonian as [in analogy with Eq.\,(\ref{DSG4})]  
\begin{equation}
\hat{\cal H}={\cal H}(\hat{N})={\cal H}(q\,\partial/\partial q) ~ 
\hat{Q} \Bigr\vert_{q \to 1}
\label{OSC5}
\end{equation}
where the limit $q \to 1$ is achieved simultaneously with $D \to \infty$.  
The deformed algebra defined by the elements 
$\hat{A}, \hat{A}^{\dagger}$ and $\hat{Q}$ is an admissible  version of 
the well-known (deformed) q-oscillator algebra naturally admitting real and 
non-negative norms in the finite dimensional cyclic Hilbert space $H_{D}$.  
Since the deformation parameter 
$q=e^{-i\gamma_0{\vec m}\times{\vec m}^{\prime}}$ is a pure phase with the 
property that $q^{D}=1$, the operators $\hat{A},\hat{A}^{\dagger},\hat{Q}$
act on in the finite-$D$-dimensional cyclic Fock space 
spanned by the cyclic orthonormal vectors $\{\vert n\rangle\}=
\{\vert n\rangle_{0 \le n \le (D-1)};~~\vert n\rangle=\vert n+D\rangle\}$ 
with $\langle n^{\prime}\vert n\rangle=\delta_{n^{\prime},n}$ as,
\begin{equation}
\begin{array}{rl}
\hat{A}\,\vert n\rangle=&\sqrt{f(n)}\,\vert n-1\rangle \\
\hat{A}^{\dagger}\,\vert n\rangle=&\sqrt{f(n+1)}\,\vert n+1\rangle \\
\hat{Q}\,\vert n\rangle=&q^{n}\,\vert n\rangle ~, \qquad 
q=e^{-i\gamma_0{\vec m}\times{\vec m}^{\prime}}
\end{array}
\label{OSC6}
\end{equation}
with $0 \le f(n)$ and $f(n)=f(n+D)$ where  
\begin{equation}
f(n)=\frac{q^{n+(D-1)/2}-q^{-n-(D-1)/2}}{q-q^{-1}}+C~,
\qquad
C=\frac{2}{\vert q-q^{-1}\vert} \ne 0~. 
\label{OSC7}
\end{equation}
The algebra in Eq's\,(A.4) and the relations (\ref{OSC6}) admit a unitary 
canonical partner to $\hat{Q}$, i.e. the unitary quantum phase operator 
$\hat{\cal E}_{\phi}$
\begin{equation}
\hat{\cal E}_{\phi}=\sum_{n=0}^{D-1}\,\vert n-1\rangle \langle n\vert~,
\qquad \hat{\cal E}_{\phi}^{D} \equiv \mbox{\ee I}
\label{OSC8}
\end{equation}
such that 
\begin{equation}
\hat{Q}^{\Gamma}\,\hat{\cal E}_{\phi}^{\lambda}=q^{\Gamma\,\lambda}\,
\hat{\cal E}_{\phi}^{\lambda}\,\hat{Q}^{\Gamma}~, \qquad \Gamma, \lambda \in 
\mbox{\ee R}~. 
\label{OSC8.b}
\end{equation}
The eigenvectors of $\hat{\cal E}_{\phi}$ are $\{\vert \phi\rangle\}=
\Bigl\{\phi\rangle_{0\le r\le (D-1)}; \vert \phi\rangle_{r+D} \equiv 
\vert \phi\rangle_{r}\Bigr\} \in H_{D}$ with 
$_{r^\prime}\langle \phi\vert \phi \rangle_{r}=\delta_{r^{\prime},r}$ where 
\begin{equation}
\hat{\cal E}_{\phi}\,\vert \phi \rangle_{r}=
e^{i\,\gamma_{0} r}\,\vert \phi \rangle_{r} \qquad
\vert \phi \rangle_{r}=\frac{1}{\sqrt{D}}\,
\sum_{n=0}^{D-1}\,e^{i\,\gamma_{0} n r}\,\vert n \rangle~\qquad 
\gamma_0=\frac{2\pi}{D}~.  
\label{OSC9}
\end{equation}
The dynamical time evolution of $\hat{\cal E}_{\phi}$ for the generalized 
Hamiltonian in Eq.\,(\ref{OSC5}) is given by 
Eq.\,(\ref{DSG6.b}) as  
\begin{equation}
\begin{array}{rl}
i \frac{d}{dt}\,\langle n^{\prime} \vert \, \hat{\cal E}_{\phi}^{r}\,
\vert n \rangle=&
-\langle n^{\prime} \vert \, \hat{\cal E}_{\phi}^{r}\, 
\vert n \rangle\,\Bigl\{{\cal H}(q\,\partial/\partial q)~ 
(q^{n+r}-q^{n})\Bigl\vert_{q \to 1}\,\Bigr\} \\ 
=&-\langle n^{\prime} \vert \, \hat{\cal E}_{\phi}^{r}\,
\vert n \rangle\,\Bigl\{{\cal H}(n+r)-{\cal H}(n)\Bigr\} 
\end{array}
\label{OSC10}
\end{equation}
and a close inspection of Eq.\,(\ref{OSC10}) with (\ref{DSG6.b}) indicates 
that the pair $(\hat{Q},\hat{\cal E}_{\phi})$ is the corresponding 
{\it unitary~action-angle~pair} for the Hamiltonian in Eq.\,(\ref{OSC5}). As  
mentioned before, the $q \to 1$ and the $D \to \infty$ limits are to be taken
simultaneously on both sides of Eq.\,(\ref{OSC10}). In this limit the 
phase operator $\hat{\cal E}_{\phi}$ is the {\it unitary~version} of the 
Hermitian Pegg-Barnett phase operator\cite{29}. Making use of the fact that 
$\hat{Q}=q^{\hat{N}}$, Eq.\,(\ref{OSC8.b}) 
yields the Susskind-Glogower-Carruthers-Nieto commutation 
relations\cite{5,7,15} for the 
operator pair $(\hat{N}, \hat{\cal E}_{\phi})$. 
Note that, the QHO described by ${\cal H}=\omega\,n$ yields the equation 
of motion for the phase operator in Eq.\,(\ref{OSC10})  
which is formally identical to the equation of motion of the canonical 
angle variable for the classical harmonic oscillator. When we examine the 
quantum action-angle formalism of the generalized quantum oscillator in 
section IV.B, the 
harmonic oscillator will be realized in the specific limit when the dimension 
of the discrete phase space representations of the generalized oscillator is 
extended to infinity. 

\section{An equivalent realization of the Wigner function by  
$(\hat{\cal G}, \hat{\cal O})$} 
\subsection{A generalized approach to the AA Wigner 
function using continuous phase space representations}
It was shown that the generalized canonical phase space representation of a 
quantum 
system can be given based on the duality between the discrete Wigner-Kirkwood 
and the unitary cyclic Schwinger operator bases\cite{15,16} in 
Eq.\,(\ref{WK1}). An alternative 
to this approach is to formulate the same problem using   
the canonical transformation $\hat{\cal G}$ and its unitary canonical 
partner $\hat{\cal O}$.   

The properties of $(\hat{\cal G}, \hat{\cal O})$ studied in sections 
II and III manifest a full analogy to those of  
$(\hat{\cal U}, \hat{\cal V})$ in Eq's\,(\ref{WK2.1}-\ref{WK2.3}).  
Among the four equivalent choices in (\ref{WK2.3}), we define this analogy 
by the correspondence  
\begin{equation}
{\hat{\cal U} \choose \hat{\cal V}} \Leftrightarrow 
{\hat{\cal O} \choose \hat{\cal G}}~. 
\label{RS1}
\end{equation}

It is now suggestive to define a Schwinger operator basis labeled by 
${\vec \tau}=(\tau_1,\tau_2) \in \mbox{\ee R} \times \mbox{\ee R}$ and defined as   
\begin{equation}
\hat{\Sigma}_{\vec \tau}=e^{-i\tau_{1}\tau_{2}/2}\,\hat{\cal O}^{\tau_{1}}\,
\hat{\cal G}^{\tau_{2}}=
e^{i\tau_{1}\tau_{2}/2}\,\hat{\cal G}^{\tau_{2}}\,\hat{\cal O}^{\tau_{1}}~, 
\label{RS2}
\end{equation}
Before we study the algebraic properties of $\hat{\Sigma}_{\vec \tau}$, we 
look into some of the tracial properties of $\hat{\cal G}$ and $\hat{\cal O}$ 
operators. Since we consider the continuum limit, it is more appropriate 
to examine $\hat{\cal G}$ and $\hat{\cal O}$ in their continuous 
representation. Starting with Eq's\,(\ref{CT29}) we choose the 
$\vert h^{\chi}(\gamma)\rangle$ basis for their representation as  
\begin{equation}
\hat{\cal O}^{\tau_1}=\int_{-\infty}^{\infty}\,d\gamma\,
\vert h^{\chi}(\gamma+\tau_1)\rangle \langle h^{\chi}(\gamma)\vert ~, \qquad 
\hat{\cal G}^{\tau_2}=\int_{-\infty}^{\infty}\,d\gamma\,e^{-i\tau_2 \gamma}\,
\vert h^{\chi}(\gamma)\rangle \langle h^{\chi}(\gamma)\vert  
\label{RS2.a}
\end{equation}
from which we obtain
\begin{equation}
Tr\Bigl\{\hat{\cal O}^{\tau_1}\,\hat{\cal G}^{\tau_2}\Bigr\}
=\int_{-\infty}^{\infty}
\,d \gamma^{\prime} \langle h^{\chi}(\gamma^{\prime}) \vert \,
\Bigl\{\hat{\cal O}^{\tau_1}\,\hat{\cal G}^{\tau_2}\Bigr\}\,\vert 
h^{\chi}(\gamma^{\prime})\rangle=2\pi\,\delta(\tau_1)\,\delta(\tau_2) \equiv 
2\pi\,\delta({\vec \tau})~. 
\label{RS2.b}
\end{equation}
Using Eq.\,(\ref{RS2.b})   
the properties of $\hat{\Sigma}_{\vec \tau}$ can be found in manifest analogy 
with those of $\hat{S}_{\vec m}$ in Eq's\,(\ref{WK3}) as 
\begin{equation}
\begin{array}{rll}
\hat{\Sigma}_{\vec \tau}^{\dagger}=&\hat{\Sigma}_{-{\vec \tau}} \\
Tr\Bigl\{\,\hat{\Sigma}_{\vec \tau}\,\Bigr\}=&
2\pi\,\delta({\vec \tau}) \\
\hat{\Sigma}_{\vec \tau}\,\hat{\Sigma}_{\vec \tau^{\prime}}=&
e^{i\,{\vec \tau}\times{\vec \tau^{\prime}}/2}\,
\hat{\Sigma}_{{\vec \tau}+{\vec \tau^{\prime}}} \\
(\hat{\Sigma}_{\vec \tau}\,\hat{\Sigma}_{{\vec \tau}^{\prime}})\,
\hat{\Sigma}_{{\vec \tau}^{\prime \prime}}=&
\hat{\Sigma}_{\vec \tau}\,(\hat{\Sigma}_{{\vec \tau}^{\prime}}\,
\hat{\Sigma}_{{\vec \tau}^{\prime \prime}}) \qquad &{\rm (associativity)}\\
\hat{\Sigma}_{\vec 0}=&\mbox{\ee I} &{\rm (unit~element)}\\
\hat{\Sigma}_{\vec \tau} \hat{\Sigma}_{-\vec \tau}=&\mbox{\ee I} &{\rm 
(inverse)}~. 
\end{array}
\label{RS3}
\end{equation}
Hence, the canonical transformation generator $\hat{\cal G}$ and its unitary 
canonical partner $\hat{\cal O}$ form a continuous realization of Schwinger's 
operator basis. 

Using Eq\,(\ref{WK1}) and the analogy manifested by Eq\,(\ref{RS1})
we construct a dual form for the Wigner-Kirkwood operator basis using the 
realization of the Schwinger basis defined 
in Eq's\,(\ref{RS2}) and (\ref{RS3}) as 
\begin{equation}
\hat{\Delta}_{CT}({\vec V})=\int
\,\frac{d{\vec \tau}}{2\pi}\,e^{-i\,{\vec \tau}\times{\vec V}}\,
\hat{\Sigma}_{\vec \tau} ~,\qquad 
\hat{\Sigma}_{\vec \tau}=\int
\frac{d{\vec V}}{2\pi}\,
e^{i\,{\vec \tau}\times{\vec V}}\,
\hat{\Delta}_{CT}({\vec V})
\label{RS4}
\end{equation}
where the integrals are to be considered in $\mbox{\ee R}\times\mbox{\ee R}$.  
Using the properties of $\hat{\Sigma}_{\vec \tau}$ it can be shown that 
Eq's\,(\ref{RS4}) provide an operator basis for the Wigner function as
\begin{equation}
\begin{array}{rl}
\hat{\Delta}_{CT}({\vec V})=&\hat{\Delta}_{CT}^{\dagger}({\vec V}) \\
\int\,\frac{d{\vec V}}{2\pi}\,\hat{\Delta}_{CT}({\vec V})=&\mbox{\ee I} \\
Tr\Bigl\{\hat{\Delta}_{CT}({\vec V})\Bigr\}=&\mbox{\ee I} \\
Tr\Bigl\{\hat{\Delta}_{CT}({\vec V})\,
\hat{\Delta}_{CT}({\vec V}^{\prime})\Bigr\}=&
\delta({\vec V}-{\vec V}^{\prime})~. 
\end{array}
\label{RS5}
\end{equation}
The properties (\ref{RS5}) are necessary and sufficient conditions 
in order to define a correspondence between an arbitrary operator 
$\hat{F}$ and its Weyl-Wigner-Moyal (WWM) symbol $f({\vec V})$ 
\begin{equation}
\hat{F}=\int\,
d{\vec V}\,f({\vec V})\,\hat{\Delta}_{CT}({\vec V})~, \qquad 
f({\vec V})=Tr\Bigl\{\hat{F}\,\hat{\Delta}_{CT}({\vec V})\Bigr\}
\label{RS6}
\end{equation}
with the condition that $\Vert \hat{F} \Vert=\int d{\vec V} \, 
\vert f({\vec V})\vert^{2} < \infty$. A few simple examples can be given 

a) for $\hat{F}=\hat{\cal G}^{\Lambda}$

Using the same relations as in (a), the WWM symbol of 
$\hat{\cal G}^{\Lambda}$ is 
\begin{equation}
g_{\Lambda}({\vec V})=Tr\Bigr\{\hat{\cal G}^{\Lambda}\,
\hat{\Delta}_{CT}({\vec V})\Bigr\}=e^{-i\Lambda\,V_1}
\label{RS7}
\end{equation}

and similarly

b) for $\hat{F}=\hat{\cal O}^{\zeta}$ 

\begin{equation}
o_{\zeta}({\vec V})=Tr\Bigr\{\hat{\cal O}^{\zeta}\,
\hat{\Delta}_{CT}({\vec V})\Bigr\}=e^{i\zeta\,V_2}
\label{RS8}
\end{equation}

c) A particularly interesting case arises when the arbitrary operator 
$\hat{F}$ is invariant under a specific unitary transformation by 
$\hat{\Sigma}_{\vec \mu}$ such that 
\begin{equation}
\hat{\Sigma}_{-\vec \mu}\,\hat{F}\,\hat{\Sigma}_{\vec \mu}=\hat{F}~. 
\label{RS9}
\end{equation}
Such an operator has a translationally invariant WWM symbol 
\begin{equation}
f({\vec V})=f({\vec V}+{\vec \mu})
\label{RS10}
\end{equation} 
where, since $\Lambda$ is arbitrary, it is implied that $f({\vec V})$  
is independent of ${\vec V}.{\vec n}_{\mu}$ where ${\vec n}_{\mu}=
{\vec \mu}/\vert {\vec \mu}\vert$ is the unit vector in the ${\vec \mu}$ 
direction. 

For $\hat{\cal G}$ describing  
the elements of the dynamical symmetry group, $\Delta_{CT}({\vec V})$ 
corresponds to the quantum action-angle operator representation of the 
Wigner-Kirkwood basis. Based on the analogy in Eq.\,(\ref{RS1}) and 
the generalized Wigner function defined in Eq.\,(\ref{WK5}) we now define 
a {\it generalized~action-angle} Wigner function of an arbitrary quantum state 
$\vert \psi\rangle$ as 
\begin{equation}
W_{\psi}({\vec V})=\langle \psi\vert \Delta_{CT}({\vec V})\vert \psi\rangle
\label{RS11}
\end{equation}
which can be expressed in the continuous {\it action~eigenbasis} as 
\begin{equation}
W_{\psi}({\vec V})=\int_{-\infty}^{\infty}
\,\frac{d\tau_1}{2\pi}\,e^{i\tau_1\,V_2}\,
\langle \psi\vert h^{\kappa}(V_1-\tau_1/2) \rangle \,
\,\langle h^{\kappa}(V_1+\tau_1/2)\vert \psi\rangle
\label{RS12}
\end{equation}
and in the continuous {\it angle~eigenbasis} as  
\begin{equation}
W_{\psi}({\vec V})=\int_{-\infty}^{\infty}\,
\frac{d\tau_2}{2\pi}\,e^{-i\tau_2\,V_1}
\langle \psi\vert k^{\chi}(V_2-\tau_2/2)\rangle\,
\langle k^{\chi}(V_2+\tau_2/2)\vert \psi\rangle~. 
\label{RS13}
\end{equation}
Thus, Eq's\,(\ref{WK5}) and (\ref{RS11}) are two equivalent phase space 
representations of the same quantum system. The former is defined 
in a generic canonical basis $(\hat{\cal U}, \hat{\cal V})$, 
whereas the latter is expressed in terms of the elements of the dynamical 
symmetry group of the same system. In Eq's\,(\ref{RS12}) and (\ref{RS13}) 
$V_1$ and $V_2$ are, by the WWM correspondence in Eq's\,(\ref{RS7}) and 
(\ref{RS8}), the generalized classical action-angle variables.  

\subsection{AA Wigner function for the generalized oscillator with the 
discrete phase space representations} 
In the previous sections we examined the generalized theory of quantum 
action-angle formalism using the continuous representations of the generalized 
action-angle operators. Here, we particularize this formalism to that  
of the representations of a generalized oscillator with a discrete spectrum 
in the finite -D- dimensional Hilbert space $H_{D}$ by constructing 
the unitary canonical pair $(\hat{\cal G}, \hat{\cal O})$ as 
\begin{equation}
{\hat{\cal G} \choose \hat{\cal O}} \Leftrightarrow
{\hat{\cal E}_{N} \choose \hat{\cal E}_{\phi}}
\label{RS14.a}
\end{equation}
where $\hat{\cal E}_{N}=e^{-i\gamma_{0}\hat{N}}$ with $\hat{N}$ as defined 
in Eq's\,(A.3), (A.4) and (\ref{OSC6}). 
The unitary operator 
$\hat{\cal E}_{\phi}$ corresponding to the unitary angle operator 
$\hat{\cal O}$ above will be represented by the unitary quantum phase operator 
defined in Eq's\,(\ref{OSC8}), (\ref{OSC8.b}) and (\ref{OSC9}). 
To facilitate the correspondence with 
the classical case, we switch from the generalized notation 
${\vec V}=(V_1,V_2)$ of the action-angle variables in section IV.A  
to the more standard one $(J,\theta)$. 
The realization of the action angle Wigner-Kirkwood basis 
in the unitary number-phase basis $(\hat{\cal E}_{N},\hat{\cal E}_{\phi})$ 
has been derived in Ref.\,[15] for the generalized oscillator with  
discrete, cyclic and finite -D- dimensional Hilbert space 
representations as
\begin{equation}
\hat{\Delta}_{CT}(J,\theta)=\frac{1}{2\pi D}\,\sum_{\vec m}\,
e^{i(\gamma_0 m_1 J-m_2 \theta)}\,e^{-i\gamma_0 m_1 m_2/2}\,
\hat{\cal E}_{N}^{m_1}\,\hat{\cal E}_{\phi}^{m_2}~,\qquad {\vec m} \in 
\mbox{\ee Z}_{D} \times \mbox{\ee Z}_{D}~.   
\label{RS14.b}
\end{equation}
In the discrete case, the set of completeness relations analogous to the 
continuous ones in Eq.\,(\ref{RS5}) -by direct use of Eq.\,(\ref{RS14.b})- are 
\begin{equation}
\begin{array}{rl}
\hat{\Delta}_{CT}(J,\theta)=&\hat{\Delta}^{\dagger}_{CT}(J,\theta) \\
\int dJ\,\int d\theta\,\hat{\Delta}_{CT}(J,\theta)=&\mbox{\ee I} \\
Tr\Bigl\{\hat{\Delta}_{CT}(J,\theta)\Bigr\}=&1/2\pi \\
Tr\Bigl\{\hat{\Delta}_{CT}(J,\theta)\,
\hat{\Delta}_{CT}(J^{\prime},\theta^{\prime})
\Bigr\}=&1/2\pi\,\delta(J-J^{\prime})\,\delta(\theta-\theta^{\prime})~. 
\end{array}
\label{RS14.b.a}
\end{equation}
The AA Wigner function of $\hat{\Delta}_{CT}(J,\theta)$ in a physical state 
$\vert \psi\rangle$ is then given by\cite{15} 
\begin{equation}
W(J,\theta)=\langle \psi \vert \hat{\Delta}_{CT}(J,\theta) \vert \psi \rangle=
\frac{1}{2\pi}\,
\sum_{k=0}^{D-1}\,e^{-ik\theta}\,\langle \psi \vert J-k/2\rangle \, 
\langle J+k/2\vert \psi\rangle ~,\qquad 
k \in \mbox{\ee Z}
\label{RS14.c}
\end{equation}
where the states $\vert J \pm k/2\rangle$ are vectors in  
continuously shifted Fock spaces.\cite{15} Here 
$\{\vert J \pm k/2\rangle; k=odd\} \in 
{\cal F}^{(\alpha\pm 1/2)}$ and 
$\{\vert J \pm k/2\rangle; k=even\} \in
{\cal F}^{(\alpha)}$ with $\alpha$ satisfying the conditions\cite{15}  
that $2(J-\alpha) \in \mbox{\ee Z}$ and $\alpha \in \mbox{\ee R}[0,1)$. 
The definition of an arbitrary vector in the continuously shifted Fock  
space ${\cal F}^{(\beta)}$ has been given in Ref.\,[15] by
\begin{equation}  
\vert n+\beta\rangle \equiv \frac{1}{\sqrt{D}}\,
\sum_{\ell=0}^{D-1}\,e^{-i\gamma_0(n+\beta)\ell}\,\vert \phi\rangle_{\ell}
~, \qquad \beta \in \mbox{\ee R}[0,1)~, \qquad 
\vert n+\beta\rangle \equiv \vert n+D+\beta\rangle \in 
{\cal F}^{(\beta)}~.  
\label{RS15}
\end{equation}
Now, let us take $D \to \infty$ in Eq.\,(\ref{RS14.b}) 
and, using Eq.\,(\ref{RS15}) examine Eq.\,(\ref{RS14.c}) for 
$\vert \psi\rangle$ being 

a) a pure Fock state, i.e. $\vert \psi\rangle_{p}=\vert n\rangle$ ~, 
$n$ fixed 

and 

b) a typical mixed Fock state of the type 
$\vert \psi\rangle_{m}=(\vert n\rangle+\vert n-1\rangle)/\sqrt{2}$~, 
$n$ fixed~.  
\vskip1.5truecm
a) For $\vert \psi\rangle_{p}$, and after a short calculation,  
Eq.\,(\ref{RS14.c}) can be evaluated in the limit $D \to \infty$ as 
\begin{equation}
W(J,\theta)\Big\vert_{p}=\frac{1}{2\pi}\,\delta(n-J)
\label{RS16}
\end{equation}
The marginal probability distributions for the $J$ or $\theta$ variables  
in the pure Fock state can then be found by integrating over the other 
variable $\theta$ or $J$ respectively as  
\begin{equation}
\begin{array}{rl}
P(J)\Big\vert_{p} \equiv &\int\,d\theta\,W_{p}(J,\theta)=\delta(n-J)~, \\
\tilde{P}(\theta)\Big\vert_{p} \equiv &\int\,dJ\,W_{p}(J,\theta)=
1/2\pi
\end{array}
\label{RS17}
\end{equation}
which correctly describe the expected results for the pure Fock state. 

b) The state $\vert \psi\rangle_{m}$ is the so called {\it split} state.  
For this state, using Eq.\,(\ref{RS15}), we obtain 
\begin{equation}
W(J,\theta)\Big\vert_{m}=\frac{1}{4\pi}\,\Bigr\{\delta(n-J)+
2\,\delta(n-J-1/2)\,\cos{\theta}+\delta(n-J-1)\Bigr\}~. 
\label{RS18}
\end{equation}
The marginal probability distributions yield for $\vert \psi\rangle_{m}$  
\begin{equation}
\begin{array}{rl}
P(J)\Big\vert_{m}=&\Bigl\{\delta(n-J)+\delta(n-J-1)\Bigr\}/2 \\
\tilde{P}(\theta)\Big\vert_{m}=&(1+\cos{\theta})/2\pi
\end{array}
\label{RS19}
\end{equation}
which are the correct action and angle probability distributions 
for the split state.

The Eq.\,(\ref{RS14.c}) also provides the correct time dependence for the 
AA Wigner function in the QHO limit. 
In order to observe this we will start with the generalized oscillator 
in Eq.\,(\ref{OSC5}). 
The time dependence of the AA Wigner function is given by the standard 
expression 
\begin{equation}
i\frac{d}{dt}\,W_{\psi}(J,\theta)=\langle \psi\vert[\hat{\cal H},
\hat{\Delta}_{CT}(J,\theta)]\vert \psi \rangle
\label{RS20}
\end{equation}
or, equivalently, in terms of the WWM symbol $h(J,\theta)$ of $\hat{\cal H}$ 
as\cite{30}
\begin{equation}
\frac{d}{dt}\,W_{\psi}(J,\theta)=\Bigl\{h(J,\theta) * W_{\psi}(J,\theta)-
W_{\psi}(J,\theta) * h(J,\theta)\Bigr\}~, \qquad 
*=exp\Bigl\{\frac{i}{2}\Bigl[\frac{\stackrel{\gets}{\partial}}{\partial J}\,
\frac{\stackrel{\to}{\partial}}{\partial \theta}-
\frac{\stackrel{\gets}{\partial}}{\partial \theta}\,
\frac{\stackrel{\to}{\partial}}{\partial J}\Bigr]\Bigr\}
\label{RS21}
\end{equation}
where $(h*W_{\psi}-W_{\psi}*h) \equiv \Bigl\{h,W_{\psi}\Bigr\}_{MB}$ is the 
Moyal (sine) bracket.\cite{30} The calculation of Eq.\,(\ref{RS21}) 
requires the knowledge of $h(J,\theta)$. This can be   
obtained by using the completeness equations in (\ref{RS14.b.a}) 
and the Hamiltonian operator $\hat{\cal H}$ in (\ref{OSC5}) as 
\begin{equation}
\hat{\cal H}=\int\,dJ\,\int\,
d\theta\,h(J,\theta)\,\hat{\Delta}_{CT}(J,\theta)~, 
\qquad \, {\rm where} \qquad 
h(J,\theta)=Tr\Bigl\{\hat{\cal H} \hat{\Delta}_{CT}(J,\theta)\Bigr\}~. 
\label{RS22}
\end{equation}
The trace operation can be conveniently carried in the discrete finite 
dimensional cyclic eigenspace 
$\{\vert n\rangle\}=\{\vert n\rangle_{0 \le n \le (D-1)}; 
\vert n+D\rangle\equiv \vert n\rangle \}$ of the Hamiltonian $\hat{\cal H}$. 
Since for the diagonal matrix elements  
$\langle n \vert \hat{\Delta}_{CT}(J,\theta)\vert n\rangle=
W(J,\theta)\Big\vert_{p}$ 
are Wigner functions of the pure Fock states, we can also directly use the 
expression (\ref{RS16}) in the calculation of the trace. We find that 
\begin{equation}
h(J,\theta)={\cal H}(J)
\label{RS23}
\end{equation}
Using Eq.\,(\ref{RS23}) in Eq.\,(\ref{RS21}) 
\begin{equation}
\frac{d}{dt}W_{\psi}(J,\theta)=-i\,
\Bigl\{{\cal H}(J+\frac{i}{2}\frac{\partial}{\partial \theta})-
{\cal H}(J-\frac{i}{2}\frac{\partial}{\partial \theta})\Bigr\}\,
W_{\psi}(J,\theta)
\label{RS24}
\end{equation}
Eq.\,(\ref{RS24}) is the equation of motion of the action-angle Wigner function
 for an arbitrary generalized cyclic oscillator in Eq.\,(\ref{OSC5}). 
Now, we apply 
Eq.\,(\ref{RS24}) to the QHO case where the Hamiltonian in Eq.\,(\ref{OSC5}) 
is a linear operator of $\hat{N}$, let us say 
${\cal H}(\hat{N})=\omega\,\hat{N}$ 
with $\omega$ describing the oscillator frequency. Then by Eq.\,(\ref{RS22}) 
and (\ref{RS23}),   
$h(J,\theta)=\omega\,J$. For the QHO Eq.\,(\ref{RS24}) then yields, 
\begin{equation}
\Bigl\{\frac{d}{d t}-\omega\,
\frac{\partial}{\partial \theta}\Bigr\}\,W_{\psi}(J,\theta)=0~,\qquad 
\Longrightarrow \qquad  
\theta=\theta(t)=\omega\,t
\label{RS25}
\end{equation}  
namely, the time evolution of the QHO action-angle Wigner function in the phase 
space takes place on the classical manifold $(J=constant, 
\theta=\omega\,t)$ as expected. By Eq.\,(\ref{RS16}),  
The AA Wigner function for the pure Fock state is static. The full time 
dependent solution of the AA Wigner function, for instance, for the split  
state and the corresponding marginal probability distributions 
can be completely determined by inserting the solution of $\theta(t)$  
in Eq.\,(\ref{RS25}) into Eq.\,(\ref{RS18}) and (\ref{RS19}). 

\section{Conclusions}
The canonical-algebraic connection between the quantum phase problem 
and the quantum phase space has already been noticed by some other 
workers recently. 
In particular, using the generators of the angular momentum $su(2)$ algebra 
and its dual in terms of the Hermitian canonical phase operators  
Vourdas has studied\cite{31} an equivalent canonical pair to  
$(\hat{\cal G}, \hat{\cal O})$ as defined in this work. 
Our specific aim in this publication was to further the canonical algebraic 
approach   
introduced in Ref.\,[15] to unify the formulation of quantum phase with that   
of the algebraic theory of quantum canonical transformations. In this context,  
we investigated the generators of quantum canonical transformations, their 
unitary canonical partners in the Schwinger sense, as well as their 
action on the functions of canonical variables of the quantum phase space,  
in particular the Wigner function. 
Through this connection, the quantum phase is formally established as the 
unitary 
canonical partner of the quantum action operator which is also demonstrated  
for the one dimensional generalized oscillator. 

\section{acknowledgments}
The author is grateful to Dr. Laurence Barker (Bilkent University) for 
illuminating discussions particularly on the topics covered in section II.A. 
Stimulating discussion with Professor T. Dereli (Middle East Technical 
University) and Professor M. Arik (Bo\u{g}azi\c{c}i University) are also  
acknowledged. 
\newpage
\section*{Appendix}
In Ref.\,[15] we have examined two subalgebraic realizations 
of the discrete-cyclic Schwinger operator basis $\hat{\cal S}_{\vec m}$. 
In the following we will have a brief summary of them.  
Based on a fixed pair of vectors ${\vec m}, {\vec m}^{\prime}$ 
the sine algebra generated by $\hat{S}_{\vec m}$ supports two sub algebraic 
realizations.\cite{15} The standard $u_{q}sl(2)$ sub algebraic realization is 
obtained by constructing the generators
$$
\begin{array}{rl}
\hat{J}_{-} \equiv &d\,\hat{S}_{\vec m}+d^{\prime}\,
\hat{S}_{{\vec m}^{\prime}}
\\
\hat{J}_{+} \equiv &d^{*}\,\hat{S}_{-\vec m}+
d^{\prime^{*}}\,\hat{S}_{-{\vec m}^{\prime}} 
\\
\hat{L} \equiv &\hat{S}_{{\vec m}-{\vec m}^{\prime}}=
q^{\hat{J}_{3}+\frac{D}{2}}
\end{array}
\eqno{A.1}
\label{A.1}
$$
where $d\,d^{\prime^{*}}=d^{*}\,d^{\prime}=-(q^{1/2}-q^{-1/2})^{-2}$ 
and $q=e^{-i\gamma_{0}\,{\vec m} \times {\vec m}^{\prime}}$ so that
$$
\hat{J}_{\mp}\,\hat{L}=q^{\pm\,1}\,\hat{L}\,\hat{J}_{\mp}~,\qquad 
[\hat{J}_{-},\hat{J}_{+}]=-\frac{\hat{L}-\hat{L}^{-1}}
{q^{1/2}-q^{-1/2}}=-[\hat{J}_{3}+\frac{D}{2}]~. 
\eqno{A.2}
\label{A.2}
$$
This particular sub algebraic realization is sometimes referred as the 
magnetic translation group.\cite{32}  

On the other hand, more importantly for the purpose of this work, 
a second class of sub algebraic realizations exist  
in the form of an {\it admissible} q-oscillator algebra which can 
be obtained by defining 
$$
\begin{array}{rl}
\hat{A} \equiv & d\,\hat{S}_{\vec m}+d^{\prime}\,\hat{S}_{{\vec m}^{\prime}}
\\
\hat{A}^{\dagger} \equiv & d^{*}\,\hat{S}_{-\vec m}+
d^{\prime^{*}}\,\hat{S}_{-{\vec m}^{\prime}} \\
\hat{Q} \equiv & q^{-\hat{N}-(D-1)/2}=q^{1/2}\,
\hat{S}_{-({\vec m}-{\vec m}^{\prime})}
\end{array}
\eqno{A.3}
\label{A.3}
$$
where $d\,d^{\prime^{*}}=-d^{*}\,d^{\prime}=-(q-q^{-1})^{-1}$ and 
$q=e^{-i\gamma_{0}\,{\vec m} \times {\vec m}^{\prime}}$ so that
$$
\begin{array}{rl}
\hat{A}\,\hat{Q}=q^{-1}\,\hat{Q}\,\hat{A}~,\qquad 
\hat{A}^{\dagger}\,\hat{Q}=q\,\hat{Q}\,\hat{A}^{\dagger}~, \\
\hat{A}^{\dagger}\,\hat{A}=C+[\hat{N}] ~, \qquad {\rm such~that} \\
\hat{A}\,\hat{A}^{\dagger}-q\,\hat{A}^{\dagger}\,\hat{A}=(1-q)\,C+\hat{Q}
\end{array}
\eqno{A.4}
\label{A.4}
$$
where $[\hat{N}]=(\hat{Q}^{-1}-\hat{Q})/(q-q^{-1})$ and 
$C=(\vert sin(\gamma_{0} {\vec m} \times {\vec m}^{\prime}) \vert)^{-1}$. 
Eq's\,(A.4) imply that the q-oscillator spectrum is non-negative 
(i.e. $0 \le \Vert \hat{A}^{\dagger} \, \hat{A}\Vert$ where the spectrum is 
given 
by the eigenvalues of the operator $\hat{A}^{\dagger}\,\hat{A}=C+[\hat{N}]$)  
which, further implies that 
the Hilbert space is spanned by vectors with admissible (non-negative)  
norm. It was shown in Ref.\,[15] that the admissible q-oscillator algebra 
in Eq.\,(A.3) and (A.4) is crucial in establishing a   
canonical-algebraic approach to the quantum phase problem. Interested reader 
can find more detailed discussions of the admissible q-oscillator 
realizations therein. 


\begin{thebibliography}{99}
\bibitem{1} M. Born, W. Heisenberg and P. Jordan, Z. Phys. {\bf 35} (1926), 
557.  
\bibitem{2} P.A.M. Dirac, Proc. R. Soc. London A {\bf 114} (1927) 243; 
P.A.M. Dirac, 
{\it Principles of Quantum Mechanics} (Oxford University Press, London 1958). 
\bibitem{3} W. Heitler, {\it The Quantum Theory of Radiation} (Academic Press, 
 {\it The Quantum Theory of Radiation} (Academic Press, London 1958).  
\bibitem{4} W.H. Louisell, Phys. Lett. {\bf 7} (1963) 60. 
\bibitem{5} L. Susskind and J. Glogower, Physics {\bf 1} (1964) 49. 
\bibitem{6} R.J. Glauber, Phys. Rev. {\bf 130} (1963) 2529; R. J. Glauber 
Phys. Rev. {\bf 131} (1963) 2766.
\bibitem{7} P. Carruthers and M.M. Nieto, Rev. Mod. Phys. {\bf 40} (1968) 411; 
P. Carruthers and M.M. Nieto, Phys. Rev. Lett. {\bf 14} (1965) 387.  
\bibitem{8} R. Tana\'s, A. Miranowicz and Ts. Gantsog, Progress in Optics 
{\bf XXXV} (1996) 355; D.T. Pegg and S.M. Barnett, J. Mod. Opt. {\bf 44} 
(1997) 225. 
\bibitem{9} L. Mandel and E. Wolf, {\it Optical Coherence and Quantum Optics} 
(Cambridge Univ. Press 1995);
 Werner Vogel and Dirk-Gunnar Welsh, {\it Quantum Optics} 
(Akademia Verlag, Berlin 1994)  
\bibitem{10} F. Rocca and M. Siruge, Comm. Math. Phys. {\bf 34} (1973) 111.
\bibitem{11} Charles P. Boyer and Kurt Bernardo Wolf, J. Math. Phys. {\bf 16}, 
(1974) 1493; Kurt Bernardo Wolf, J. Math. Phys. {\bf 15}, (1974) 2102. 
\bibitem{12} M. Moshinsky and T.H. Seligman, Ann. Phys. (N.Y.) {\bf 114} 
(1978) 243; M. Moshinsky and T.H. Seligman, J. Phys. A {\bf 12}, (1979) L135. 
\bibitem{13} A. Luis and L.L. Sanchez-Soto, Phys. Rev. A {\bf 48} (1993) 752.
\bibitem{14} H. Ralph Lewis, Walter E. Lawrence and Joseph D. Harris, 
Phys. Rev. Lett. {\bf 77} (1996) 5157. 
\bibitem{15} T. Hakio\u{g}lu, J. Phys. A: Math. Gen. {\bf 31} (1998) 6975.
\bibitem{16} J. Schwinger, Proc. Natl. Acad. Sci. {\bf 46} (1960) 883; 
J. Schwinger, Proc. Natl. Acad. Sci. {\bf 46} (1960) 1401. 
\bibitem{17} D. Galetti and A.F.R. de Toledo Piza, Physica, {\bf A 149} (1988) 
267; R. Aldrovandi and D. Galetti, J. Math. Phys. {\bf 31} (1990), 
2987. 
\bibitem{18} E P Wigner, Phys. Rev. {\bf 40} (1932) 749; 
J G Kirkwood, Phys. Rev. {\bf 44} (1933) 31. 
\bibitem{19} D. Fairlie, P. Fletcher and C. Zachos, Phys. Lett. B {\bf 218} 
(1989) 203; D.B. Fairlie and C. Zachos, Phys. Lett. B {\bf 224} (1989) 101. 
\bibitem{20} V Arnold, Ann. Inst. Fourier {\bf XVI} (1966) 319; 
V Arnold, {\it Mathematical Methods of Classical Mechanics} (Berlin:Springer, 
1978). 
\bibitem{21} R. Balian and C. Itzykson, C.R. Acad. Sci. Paris {\bf 303},
{\bf I}, (1986) 773; G.G. Athanasiu and E.G. Floratos, Nucl. Phys. 
{\bf B425} (1994) 343.   
\bibitem{22} H. Weyl, {\it The Theory of Groups in Quantum Mechanics} 
(New York: Dover 1931). 
\bibitem{23} D. Ellinas and E.G. Floratos, {\it Prime decomposition and 
entanglement measure of finite quantum systems}, 
Los Alamos preprint quant-ph/9806007. 
\bibitem{24} Victor Namias, J. Inst. Maths. Applics {\bf 25} (1980) 241. 
\bibitem{25} Soo-Chang Pei and Min-Hung Yeh, Optics Letters {\bf 22} 
(1997) 1047. 
\bibitem{26} \c{C}a\u{g}atay Candan, M. Alper Kutay and Haldun M. Ozaktas, 
{\it The discrete fractional Fourier Transformation}, to appear in the  
{\em Proceedings of the IEEE International Conference on Acoustic, Speech and 
Signal Processing}, (1999). 
\bibitem{27} R. Gilmore, {it Lie Groups, Lie Algebras and Some of Their 
Applications}, (John Wiley, New York 1974). 
\bibitem{28} S.J. Wang and S.Y. Chu, J. Phys. A: Math. Gen. {\bf 27} (1994) 
5655.
\bibitem{29} D.T. Pegg and S.M. Barnett, Europhys. Lett. {\bf 6} (1988) 483; 
D.T. Pegg and S.M. Barnett, Phys. Rev. {\bf A 39} (1989), 1665.
\bibitem{30} J. Moyal, Proc. Camb. Phil. Soc. {\bf 45} (1949) 99.
\bibitem{31} A. Vourdas, Phys. Rev. {\bf A 41} (1990) 1653; 
A. Vourdas, J. Phys. A: Math Gen. {\bf 29} (1996) 4275; A. Vourdas, 
Phys. Rev. A {\bf 43} (1991) 1564; A. Vourdas and C. Bendjaballah, Phys. 
Rev. A {\bf 47} (1993) 3523.  
\bibitem{32} J. Zak, Phys. Rev. {\bf 134} (1964) 1602; E. Brown, Phys. Rev. 
{\bf 133} (1964) 1038; X. Shen, Int. J. Miod. Phys. {\bf A 7} (1992) 3717; 
T. Dereli and A. Ver\c{c}in, Phys. Lett. {\bf B 288} (1992) 109; 
I. Kogan, Int. J. Mod. Phys. {\bf A 9} (1994) 3887.  
\end{thebibliography}
\end{document}